\documentclass[nofootinbib,aps,prx,twocolumn,secnumarabic,balancelastpage,amsmath,amssymb,floatfix,superscriptaddress]{revtex4-1}

\usepackage{xcolor}

\usepackage{ulem}
\newcommand{\tr}{\operatorname{Tr}}
\newcommand{\ketbra}[2]{\ket {#1} \hskip -0.8ex \bra {#2}}
\newcommand{\toy}{\mathrm{toy}}
\newcommand{\expct}[1]{\left\langle #1 \right\rangle}
\newcommand{\stab}{\mathsc{stab}}

\newcommand{\diag}[2]{\vcenter{\hbox{\includegraphics[height=#1]{#2}}}}

\usepackage{amssymb}
\usepackage{amsmath}
\usepackage{amsfonts}
\usepackage{braket}
\usepackage{graphicx}
\usepackage{epstopdf}
\usepackage{epsfig}
\usepackage[toc,page,title,titletoc,header]{appendix}
\usepackage{dsfont,amsthm,amsbsy}
\usepackage{hyperref}
\usepackage[xspace]{ellipsis}
\usepackage{cancel}

\newcommand{\mathsc}[1]{\textnormal{\textsc{#1}}}

\begin{document}
\title{Conformal field theories are magical}
\author{Christopher David White}
\affiliation{Condensed Matter Theory Center, University of Maryland, College Park, Md, 20742}
\author{ChunJun Cao}
\affiliation{QuICS, University of Maryland, College Park, Md, 20742}
\author{Brian Swingle}
\affiliation{Condensed Matter Theory Center, University of Maryland, College Park, Md, 20742}

\date{\today}

\begin{abstract}
  ``Magic'' is the degree to which a state cannot be approximated by Clifford gates.
  We study mana,
  a measure of magic,
  in the ground state of the $\mathbb Z_3$ Potts model,
  and argue that it is a broadly useful diagnostic for many-body physics.
  In particular we find that the $q = 3$ ground state has large mana at the model's critical point,
  and that this mana resides in the system's correlations.
  We explain the form of the mana by a simple tensor-counting calculation based on a MERA representation of the state.
  Because mana is present at all length scales, we conclude that the conformal field theory describing the 3-state Potts model critical point is magical.
  These results control the difficulty of preparing the Potts ground state on an error-corrected quantum computer,
  and constrain tensor network models of AdS-CFT.
\end{abstract}

\maketitle


\section{Introduction}

The study of quantum entanglement has brought about meaningful progress at the intersection of physics and information theory,
by helping to characterize topological phases of matter\cite{Kitaev:2005dm, Levin2006},
understand how isolated systems thermalize
\cite{deutsch_quantum_1991,srednicki_chaos_1994,rigol_thermalization_2008,dalessio_quantum_2016}
or fail to thermalize \cite{bardarson_unbounded_2012,serbyn_universal_2013,kjall_many-body_2014}
and diagnose the structure of spacetime in quantum gravity \cite{Ryu:2006bv,Hubeny:2007xt,Swingle:2009bg,VanRaamsdonk:2010pw,Maldacena:2013xja,Qi:2013caa,Cao:2016mst,Cao:2017hrv}.
Entanglement is also a crucial resource for quantum communication and quantum computing \cite{Teleportation93,GottesmanChuang99,Nielsen2003,NielsenChuang,Horodecki:2009zz}. 
However, entanglement alone does not fully account for the power of a universal quantum computer, nor does it fully characterize the rich structure of quantum information in physical systems.

Clifford operations provide an important example of the limitations of entanglement as a diagnostic
in many-body physics.
Cliffords are are a special set of non-universal quantum operations that can be efficiently simulated on a classical computer despite being capable of generating superpositions and maximally entangled states \cite{gottesman_heisenberg_1998}.
These operations and the states they generate,
the so-called ``stabilizer states'',
have played important roles in the study of
topological order \cite{kitaev_fault-tolerant_2003},
thermalization \cite{gopalakrishnan_facilitated_2018,choi_quantum_2019},
and quantum gravity \cite{pastawski_holographic_2015}
but there too they fail to capture the full richness of the physics.

Clifford operations also have a special role in quantum computation.
The Eastin-Knill theorem \cite{eastin_restrictions_2009} states that no single error-correcting code can implement universal quantum computation transversally.
Many gates may be ``easy'' in a particular code,
in the sense that they can be implemented without coupling subsystems
and so are immediately fault-tolerant;
but some will be subject to error.
Frequently the Clifford gates are ``easy'' in this sense, and the non-Clifford gate, \textit{e.g.} the T gate, is hard.
This is true for many stabilizer codes;
it is also true for Majorana-based topological quantum computation
(but see \cite{karzig_optimal_2015,karzig_geometric_2016,karzig_robust_2018,brown_fault-tolerant_2020}). 

We propose that non-Cliffordness---or ``magic'', as it is known in the quantum information community---is an important physical property that, like entanglement, sheds light on a variety of problems in quantum many-body physics and quantum simulation of field theories.
We focus on magic in the ground states of conformal field theories.
This is relevant, for instance, in the context of state preparation and quantum gravity.

The difficulty of simulating quantum systems on classical computers has motivated work on quantum computers since their inception \cite{feynman_simulating_1982}.
Jordan, Lee, and Preskill \cite{jordan_quantum_2012} showed that a quantum computer could extract the dynamics of a (relativistic) quantum field theory with polynomial resources,
by preparing a ground state,
creating excitations on top of that ground state,
and simulating the dynamics of those excitations.
But gate count simpliciter is not adequate to estimate performance,
because T gate applications will likely be the most resource-intensive operations.
The magic of a field theory ground state therefore controls the difficulty of the first step---state preparation---in quantum simulation.

At the same time, we argue that magic monotones probe the information structure of those ground states in a way
that sheds light on the AdS-CFT correspondence.
The Ryu-Takayanagi formula \cite{Ryu:2006bv} motivates tensor network toy models, which frequently consist of ``AdS/MERA''~\cite{Swingle:2009bg, Swingle:2012wq}---a so-called ``multiscale entanglement renormalization ansatz'' tensor network~\cite{Vidal2008} in the bulk, producing a state with logarithmic entanglement in the boundary---or of carefully arranged quantum error correcting codes \cite{pastawski_holographic_2015,jahn_tensor_2020}; cf \cite{cui_bit_2019}
(although other ans\"atze have also seen use~\cite{Evenbly:2017hyg}).
All of these toy models give insight into the correspondence,
but none is completely satisfactory:
MERAs are underconstrained, in that they can reproduce a very wide variety of states with appropriate tensor choices,
while error-correcting codes are dramatically over-constrained,
in that they can only reproduce stabilizer states\footnote{It is possible to produce non-stabilizer states in \cite{pastawski_holographic_2015} by tuning the logical information, but it is insufficient to reproduce a CFT ground state. The random tensor network model \cite{Hayden:2016cfa} is not a stabilizer code, but it does produce a flat entanglement spectrum which is not what we expect for CFTs.}
(even worse---typically stabilizer states in which local expectation values are identically zero).

In this work we compute a certain function ``mana'', which measures magic, for ground states of the 3-state Potts model.
We find that it is nonlocal and peaked at the model's phase transition,
and we give a MERA picture explaining its origin and its dependence on the parameters we vary.
This MERA picture requires that non-Clifford rotations be part of the structure of the state at all scales.
We thus address both the difficulty of preparing ground states of quantum field theories,
and the constraints that information structure places on tensor network models of the AdS-CFT correspondence.
Crucially, both our numerical results and our MERA picture indicate that
the non-Clifford, ``magical'' structure of the theory persists into the infrared limit.
Magic in a field theory is not a lattice-scale effect,
nor an artifact of the particular theory we consider.
Rather, it is a feature of the infrared structure of the conformal field theory itself. 

In Sec.~\ref{s:magic} we briefly review magic, including the magic monotone ``mana'' we use to quantify it, and
in Sec.~\ref{s:model} we discuss the model and our numerical methods.
In Sec.~\ref{s:results}, the empirical core of the work, we discuss our numerical results.
In Sec.~\ref{s:mera} we show that the form of our results is consistent with a simple MERA picture for the system's magic, and that our results constrain the MERA tensors.
In Sec.~\ref{s:meanfield} we discuss magic in a mean-field model of the $q\ge 5$-state Potts model.
(Taken together, these two sections---on MERAs and mean-field transitions---show the extent to which our results are generic.)
In Sec.~\ref{s:discussion} we describe the implications of our results both for quantum simulation of field theories and for AdS-CFT, and give a broader perspective on other possible uses of magic in future work.

Note: One related work was presented recently in \cite{sarkar_characterization_2020}; they study magic in free-fermion systems.\footnote{Their expanded work
  (v2 of the arXiv posting \cite{sarkar_characterization_2020})
    has substantial overlap with our work. The expansion appeared after the main results of this work had been obtained and presented;
    we became aware of it during final preparation of this manuscript.}

\section{Magic}\label{s:magic}
We plan to argue that critical ground states are difficult to prepare.
To say this meaningfully, we need to distinguish easy operations from hard operations,
and articulate an operational measure of how many hard operations are required to prepare such states.

The long-term goal of quantum computation is to construct and use a large-scale fault tolerant computer.
Clifford operations are important elements in many schemes for doing this,
in part because they can often be implemented transversally---that is, without spreading single-qubit errors.
To implement universal quantum computations one only needs a simple set of Clifford gates, which can generate any Clifford operation, and one additional non-Clifford element that renders the whole set universal.
We will call the additonal non-Clifford element a magic gate.
One standard choice is a single-qubit phase gate called a T gate.
No quantum error-correction code can implement a universal set of gates transversally \cite{eastin_restrictions_2009};
the magic gates, such as T gates, are often much harder to implement fault-tolerantly.

One can circumvent the difficulty of applying magic gates by implementing them using Clifford gates and measurements applied to special input ``magic'' states.
But typically the available magic states are themselves imperfect,
so they must be ``distilled'' into nearly ideal magic states \cite{bravyi_universal_2005}.
The cost of running the computation is then often dominated by the cost of distilling magic states,
which is effectively the cost of applying non-Clifford operations.

When non-Clifford gates indeed require much more effort to implement compared to Clifford gates,
magic gates or equivalently magic states can be usefully regarded as an expensive resource for quantum computation.
So if one is interested in preparing non-trivial states on a fault tolerant quantum computer,
it is useful to characterize the magic required.
By contrast, generating lots of entanglement may be regarded as straightforward.
Then reducing the difficulty of the preparation means reducing the number of T gates
(or more generally, magic gates) in the circuit \cite{Kliuchnikov2013,Selinger2013,Gosset2014}.

Careful circuit construction \cite{gosset_algorithm_2013,amy_meet---middle_2013,amy_polynomial-time_2014,nam_automated_2018,heyfron_efficient_2018,wang_resource-optimized_2020,shaw_quantum_2020} 
will reduce T gate count. 
But how far can such optimizing compilers go?
A good estimate will be circuit-independent and provide a lower bound
in the same way that entanglement measures bound the number of non-LOCC gates required.
``Magic monotones'' \cite{veitch_resource_2014} (see \cite{howard_application_2016,wang_efficiently_2018,regula_convex_2018,bravyi_simulation_2019,beverland_lower_2019} for more monotones),
which are nonincreasing under Clifford gates,
measurement of Pauli strings,
and certain other reasonable operations,
fit the bill.
They can provide a circuit-independent estimate of the number of magic gates required to create some target state,
and hence the difficulty of preparing that state.

\subsection{Clifford gates, T gates, and stabilizer states}

We consider systems composed of tensor products of $q$-level systems. Because the theory of Clifford gates and stabilizer states is simplest when the local dimension $q$ is an odd prime, we restrict to this case. Starting from the clock and shift operators,
\begin{equation}
  \label{eq:clock-and-shift}
  Z = \sum_{n = 0}^{q-1} \omega^n \ketbra{n}{n}, \quad X = \sum_{n=0}^{q-1} \ketbra{n + 1 \mod q}{n}\;
\end{equation}
with $\omega = e^{2\pi i/q}$, generalized Pauli operators are defined as
\begin{equation}
  T_{aa'} = \omega^{-2^{-1} aa'} Z^a X^{a'}
\end{equation}
where $2^{-1} = \frac{q+1}{2}$ is the multiplicative inverse of $2 \mod q$;
Pauli strings are
\begin{equation}
  T_{\mathbf a} = T_{a_1a_1'} T_{a_2a_2'} \dots T_{a_Na_N'}\;.
\end{equation}

The \textbf{Clifford group} is the group of unitaries generated by Clifford gates;
it is also consists of exactly those unitaries
that map (generalized) Pauli strings to (generalized) Pauli strings,
up to a phase:
\begin{align}
  \mathcal C = \{U\ : \ U T_{\mathbf a} U^\dagger = e^{i\phi} T_{\mathbf b} \}.
\end{align}

The bulk of this paper studies the $q=3$ case, so for illustrative purposes we focus on $q=3$ (qutrits) in the present section. However, we emphasize that the broad formalism generalizes to any odd prime $q$. The qutrit Clifford group is generated by the phase gate, qutrit Hadamard, and sum gate
\begin{align}
  &K = \begin{bmatrix} 1 \\ &1 \\ && \omega \end{bmatrix},\ 
  H = \begin{bmatrix} 1  & 1 & 1\\ 1&\omega &\omega^2\\ 1&\omega^2& \omega \end{bmatrix},\ \\
  &\quad\qquad S = \sum_{ij} \ketbra{i;i \oplus j}{i;j}\;.
\end{align}
Because the Cliffords map each Pauli string to a single other Pauli string,
Clifford unitaries map computational basis states like $\ket{0 \dots 0}$ to any eigenstate of Pauli strings:
\begin{equation}
  \mathcal C \ket{0 \dots 0} = \{\text{eigenstates of $T_{a_1a_1'}T_{a_2a_2'} \dots T_{a_N a_N'}$}\}\;.
\end{equation}
These eigenstates are called \textbf{stabilizer states};
the stabilizer states are a proper subset of all possible quantum states.
If we allow classical randomness
(or, equivalently, circuit choice conditional upon measurement outcome),
the Cliffords can additionally reach statistical mixtures of stabilizer states---that is,
the convex hull of stabilizer states
\begin{align}
  \begin{split}
    &\textsc{stab}[(\mathbb C^3)^{\otimes N}]\\
    &\quad = \left\{ \sum_j p_j \ketbra{s}{s} : \ket s \in \mathcal C \ket{0\dots 0},\right.\\
    &\qquad\qquad\qquad \left. p_j \ge 0, \sum_j p_j = 1 \right\}\;.
  \end{split}
\end{align}
(Since we typically work in $(\mathbb C^3)^{\otimes N}$, we will call this set $\textsc{stab}$;
occasionally, when working on a different number of qutrits,
we will use $\mathsc{stab}[n]$ for $\mathsc{stab}[(\mathbb C^3)^{\otimes n}]$.)

This immediately implies that the Clifford gates are not universal:
if one starts from a computational basis state, many states are unreachable by Clifford circuits.
Furthermore, the Gottesman-Knill theorem~\cite{gottesman_heisenberg_1998} indicates that Clifford circuits can be simulated with polynomial time with a probabilistic classical computer.
To create a universal gate set we must add a non-Clifford gate.
One suitable gate is called a T gate.
For qutrits, the T gate is defined as
\begin{equation}
  T = \begin{bmatrix} \xi^{-1} \\ & 1 \\ && \xi \end{bmatrix}
\end{equation}
where $\xi = e^{2\pi i/9}$.
This is our ``hard'' operation.

\subsection{Magic monotones, discrete Wigner functions, and mana}

T gates (or in fact arbitrary rotations) can be implemented
by applying Clifford gates and measurements to special ``magic'' states.
Any function that is non-increasing under
Clifford gates,
projective measurement of Pauli operators,
and certain other sensible operations,
can therefore provide an estimate of T gate requirements.
Such a function is called a ``magic monotone'' \cite{veitch_resource_2014}.

The mana of a state $\rho$ is a relatively easy-to-compute magic monotone.
To define mana,
first define the \textbf{phase space point operators}
\begin{equation}
  A_{\mathbf b} = 3^{-N}T_{\mathbf b} \left[\sum_{\mathbf a} T_{\mathbf a}\right] T_{\mathbf b}^\dagger\;.
\end{equation}
These operators provide a Hermitian and Frobenius-orthogonal basis for $\mathbb C^{3^N \otimes 3^N}$:
\begin{equation}
  \tr A_{\mathbf b} A_{\mathbf c} = 3^N\delta_{\mathbf b \mathbf c}\;;
\end{equation}
additionally
\begin{equation}
  3^{-N}\sum_{\mathbf b} A_{\mathbf b} = I\;.
\end{equation}
Now expand the density matrix $\rho$ in this basis:
\begin{equation}
  \rho = \sum_{\mathbf u} W_{\rho}(\mathbf u) A_{\mathbf u}\;.
\end{equation}
The coefficients $W_\rho(\mathbf u)$ are collectively called the \textbf{discrete Wigner function} of $\rho$.
Because $\rho$ and the $A_{\mathbf{u}}$ are all Hermitian,
they are real: $W_\rho(\mathbf u) \in \mathbb R$.
Immediately
\begin{equation}
  \sum_{\mathbf u} W_{\rho}(\mathbf u) = \tr \rho = 1\;;
\end{equation}
additionally, the Frobenius inner product is
\begin{equation}
  \tr(\rho_1 \rho_2) = 3^N \sum_{\mathbf u} W_{\rho_1}(\mathbf u)W_{\rho_2}(\mathbf u)\;.
\end{equation}

The \textit{discrete Hudson's theorem} \cite{gross_hudsons_2006,gross_non-negative_2007,hudson_when_1974} states that
a pure state's Wigner function is positive if and only if that state is a stabilizer state:
\begin{equation}
  W_{\ket{\psi}\hskip-0.35ex\bra{\psi}}(\mathbf u) \ge 0\quad \Longleftrightarrow\quad \rho \in \mathcal C \ket{0\dots 0}\;.
\end{equation}
(For mixed states there are states with positive representation that lie outside the convex hull $\mathsc{stab}$: see \cite{gross_hudsons_2006} Sec.~5 for a concrete example.)
For a pure or mixed state $\rho$, the quantity
\begin{equation}
  \label{eq:mana}
  \mathcal M(\rho) \equiv \log \sum_{\mathbf u} | W_\rho(\mathbf u) |\;,
\end{equation}
called the \textbf{mana},
measures the magnitude of the negative components of the Wigner function.
The mana is only defined for odd onsite Hilbert space dimension.

The mana is a sensible measure of the number of non-Clifford gates required to construct $\rho$,
despite the margin between
the set of states with positive Wigner representation and
the convex hull $\mathsc{stab}$.
Mana has the additional useful properties that 
\begin{equation}
  \label{eq:product-add}
  \mathcal M(\rho_A \otimes \rho_B) = \mathcal M(\rho_A) + \mathcal M(\rho_B)\;
\end{equation}
and
\begin{equation}
  \label{eq:jensen-bound}
  \mathcal M(\rho) \le \frac 1 2 (N \log 3 - S_2)\;,
\end{equation}
where $S_2$ is the second R\'enyi entropy of $\rho$.
(The latter follows from Jensen's inequality).
Mana also measures the difficulty of the sign problem in certain classical Monte Carlo simulations \cite{pashayan_estimating_2015}.
In the Monte Carlo community it is called the ``average sign''
(or ``quenched partition function function'':
in our case the physical partition function is $\sum_{\mathbf u} W_\rho(\mathbf u) = 1$,
so the two are equivalent).

\subsection{Mana as a lower bound on number of non-Clifford rotations}\label{s:mana-bound}
Property \eqref{eq:product-add} means that the mana of a state $\rho$ lower bounds the number of non-Clifford rotations required to prepare $\rho$ from a computational basis state.
(This is essentially Theorem 14 of \cite{veitch_resource_2014}).

To see this, imagine implementing each non-Clifford rotation via magic state injection, \`a la \cite{bravyi_universal_2005}.
Magic state injection gives a protocol consisting of Clifford gates and Pauli measurements that, when applied to an input state and some number $n$ of ``magic state'' ancillae $\ket{a}$, will implement the non-Clifford rotation on the input state with success probability $p$.
So a circuit requiring $Q$ non-Clifford rotations can be implemented by Clifford operations and measurements applied to an input state
\begin{equation}
  \label{eq:msi-input}
  \ket{0\dots 0}\otimes \ket a^{\otimes (nQp^{-Q})} \approx \ket{0\dots 0}\otimes \ket a^{\otimes (nQ)}\;.
\end{equation}
The factor $p^{-Q}$ in the exponent comes from the requirement that one retry on failure.
But $p$ can be made as near 1 as one likes by taking $n$ large:
essentially, one can keep trying each magic state injection until it succeeds,
rather than running the whole circuit, noting success or failure, and re-running until success.

Suppose now that the circuit maps $\ket {0\dots 0}$ to $\rho$.
Implementing the circuit by magic state injection with input state \eqref{eq:msi-input}
and using that mana is non-increasing under Clifford operations and measurements, we find
\begin{align}
  \label{eq:mana-bound}
  \begin{split}
    \mathcal M(\rho) &\ge \mathcal M\left(\ket{0\dots 0}\otimes \ket a^{\otimes (nQ)}\right)\\
    &= nQ \mathcal M(\ket a)\;.
  \end{split}
\end{align}

\section{Model and method}\label{s:model}
\subsection{Model}
Our chosen measure of magic (Sec.~\ref{s:magic}) requires that we work with odd onsite Hilbert space dimension.
We choose the $q = 3$-state vector Potts model
\begin{equation}
  \label{eq:ham}
  H = -\sin \theta \sum_j [Z_j^\dagger Z_{j+1} + h.c.] - \cos \theta \sum_j [X^\dagger_j + X_j]
\end{equation}
where, again, $Z$ and $X$ are the clock and shift operators for $q=3$,
\begin{equation}
  Z = \sum_{n = 0}^{2} \omega^n \ketbra{n}{n}, \quad X = \sum_{n=0}^{2} \ketbra{n + 1 \mod 3}{n}\;
\end{equation}
with $\omega = e^{2\pi i/3}$.
The model has a domain wall duality transformation $\theta \leftrightarrow \pi/2 - \theta$ and a continuous phase transition at the self-dual point $\theta_c = \pi/4$;
the order parameter is (up to factors)
\begin{equation}
  m = \langle Z + Z^\dagger \rangle\;.
\end{equation}
In the ordered phase it has a $\mathbb Z_3$ symmetry.
At the critical point the model is described by a conformal field theory with central charge $4/5$ in the thermodynamic limit\cite{Cardy:1986ie,Gehlen_1986}.

The generalized Pauli operators for $q=3$ are
\begin{equation}
  T_{aa'} = \omega^{-2^{-1} aa'} Z^a X^{a'}
\end{equation}
where $2^{-1} = 2$ is the multiplicative inverse of $2 \mod 3$;
Pauli strings are
\begin{equation}
  T_{\mathbf a} = T_{a_1a_1'} T_{a_2a_2'} \dots T_{a_Na_N'}\;.
\end{equation}

As indicated in the introduction, we find that the mana of this model's ground state is maximal at the phase transition, $\theta=\pi/4$. In the limit where either $\cos \theta $ or $\sin \theta$ is zero, the ground state is a stabilizer state and hence has zero mana. Intuitively, the critical point is where the two terms, which separately favor complementary stabilizer ground states, maximally compete. Hence, it is physically reasonable that the mana is large at the critical point.

\subsection{Method}\label{s:method}
We use DMRG to find the ground state of the model \eqref{eq:ham}.
For $\theta \lesssim \pi/4$ this is straightforward.
For $\theta \gtrsim \pi/4$ the system has a three-fold degeneracy generated by $\prod_{j = 1}^N X_j$.
We first construct a ground state $\ket{\Omega_0}$ that breaks this symmetry
by biasing the DMRG initial state.
We then construct the cat ground state
\begin{equation}
  \ket \Omega = \sum_n \Big[\prod_j X_j \Big]^n \ket {\Omega_0}\;;
\end{equation}
we use this ground state for further calculations.

For $\theta \approx \pi/4$ the system is near the critical point,
and it is not obvious that the ground state has a reasonable MPS representation,
nor---if it has one---that DMRG will find that MPS.
But the bipartite entanglement entropy of the critical ground state is logarithmic in size system size\cite{calabrese_entanglement_2009},
and the gap will be polynomial in system size.
In App.~\ref{app:gs} we give some details of the ground-state calculation.
We check that our MPS ground states have logarithmic entanglement scaling for $N = 8 \dots 128$.

\section{Numerical results}\label{s:results}

\begin{figure}[t]
  \includegraphics[width=0.45\textwidth]{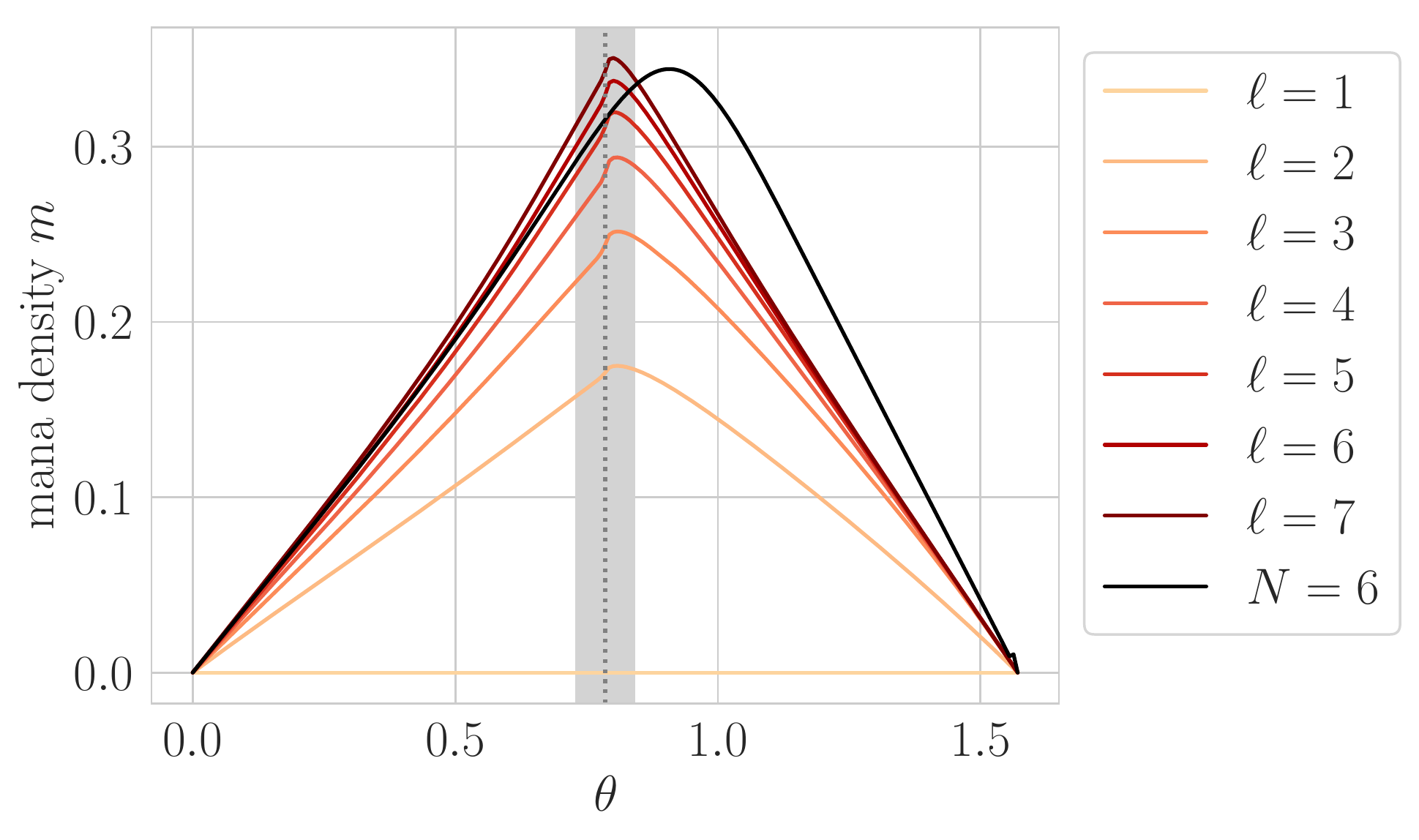}
  \caption{\textbf{Ground state mana density} for $N = 6$-site systems (black) and $l$-site subsystems of an $N = 128$-site system (orange/red).
    For $N = 128$ the mana density is peaked at the phase transition
    and appears to take the form
    $m \propto |\theta - \theta_c|$.
    Comparison with the $N = 6$ mana density
    shows that the subsystem mana is a good estimator of the whole-system mana density;
    the peak of the $N = 6$ mana density is rounded
    and shifted away from the true critical point
    due to finite-size effects.
    The light-grey region shows where the correlation length is $2\xi > 7$,
    where finite-subsystem effects become important
    and our results substantially underestimate the mana density.
  }
  \label{fig:contiguous}
\end{figure}

\subsection{Mana of simply connected subsystems}\label{ss:results:simply-connected}

Property \eqref{eq:jensen-bound} of the mana implies that the it is at most extensive;
property \eqref{eq:product-add} strongly suggests that states with short-range correlations
will in fact have extensive mana.
We therefore measure the \textbf{mana density} of $\ell$-site subsystems
\begin{equation}
  m(\rho) = \mathcal M(\rho) / \ell\;.
\end{equation}
The questions, then, are---is
the mana extensive or subextensive
at the critical point,
where the system does not have short-range correlations?
And how does the mana density vary away from the critical point?

Fig.~\ref{fig:contiguous} shows the mana density of $\ell$-site subsystems of the ground state for $\ell = 1 \dots 7$, and the mana density of the entire ground state of an $N = 6$ system.
For $\ell = 1$, the mana density is $m_{\ell = 1} (\theta) = 0$.
This is because for any one-site reduced density matrix $\rho_1$ of the ground state, $[\rho_1, X] = 0$ by construction (cf Sec.~\ref{s:method}).
$\rho$ is therefore diagonal in an $X$ eigenbasis, i.e. it is a statistical mixture of $X$ eigenstates.

For $\ell > 1$ the mana density increases with $\ell$:
the mana of this state resides in correlations.
It appears to converge, at least for away from the critical point.

We estimate finite-subsystem effects and scaling in $\theta - \theta_c$
in a MERA picture of the critical point
(Secs~\ref{ss:mera:finite} and \ref{ss:mera:nearcritical} below).
For $|\theta - \theta_c| \gtrsim 0.07\pi/4$,
the system has correlation length $2\xi \lesssim 7$,
(App.~\ref{app:corr})
so the finite-subsystem mana densities of Fig.~\ref{fig:contiguous}
accurately reflect the whole-system mana density.
For $|\theta - \theta_c| \lesssim 0.07\pi/4$,
finite-subsystem corrections become important,
and the results of Fig.~\ref{fig:contiguous} substantially underestimate the whole-system mana density.
We shade this region in light grey in the plot.

The toy model of Sec.~\ref{ss:results:twopoint} makes somewhat more concrete our notion of ``resides in the correlations''.

\subsection{Two-point mana: sudden death of magic}\label{ss:results:twopoint}

\begin{figure}[t]
  \begin{minipage}{0.45\textwidth}
    \includegraphics[width=\textwidth]{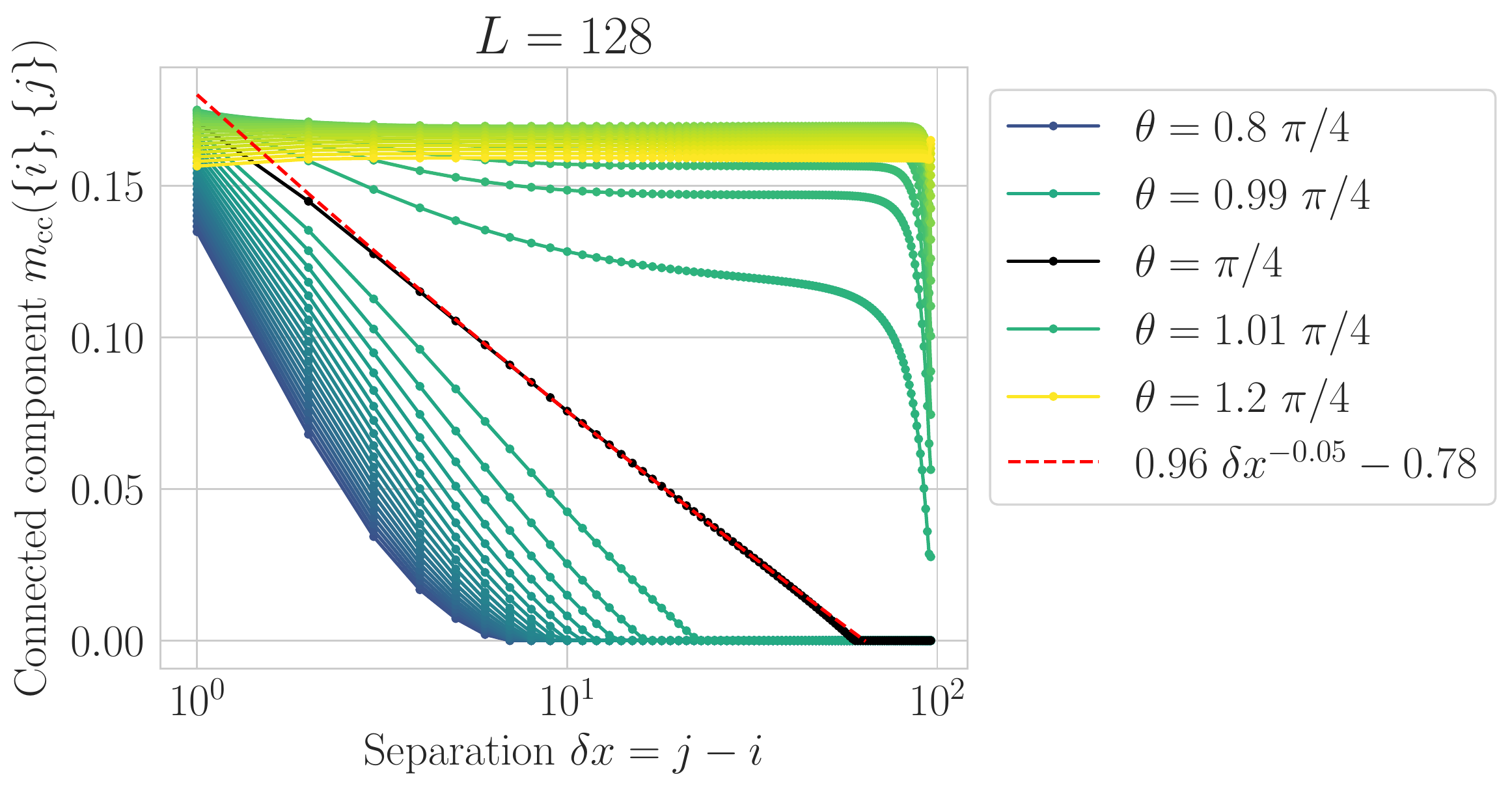}
  \end{minipage}
  
  \begin{minipage}{0.45\textwidth}
    \includegraphics[width=\textwidth]{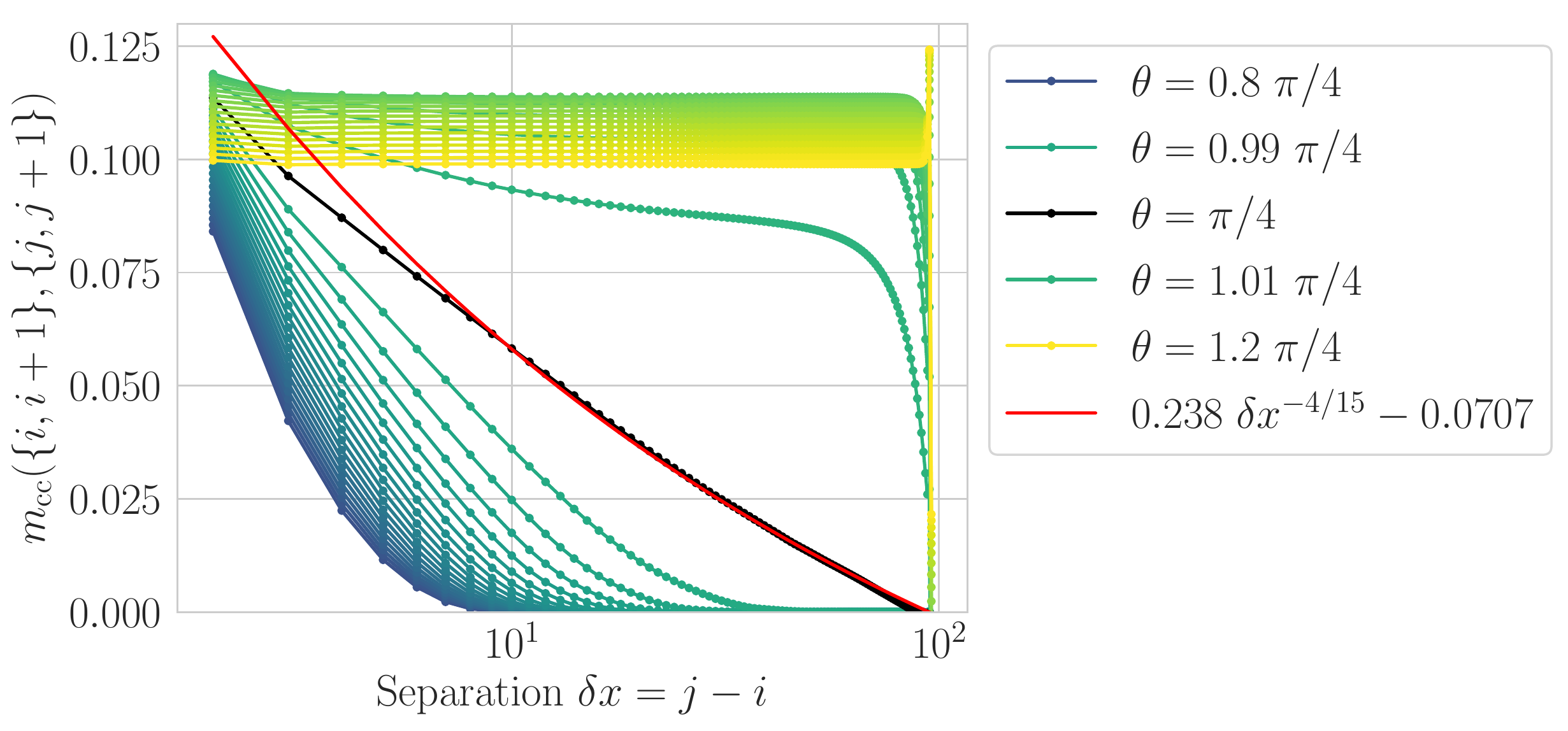}
  \end{minipage}
  \caption{\textbf{Two-point mana:} connected component of mana density $m_{\mathrm{cc}}(A,B)$ (Eq.~\ref{eq:mcc-def})
    as a function of the distance $\delta x$ between subsystems $A$ and $B$
    for $A,B$ each one site (\textbf{top}) or two sites (\textbf{bottom}).
    We take $A$ at site $32$ and $B$ right of $A$.
    The black curve is mana density at the critical $\theta = \pi/4$;
    the two red lines are two different crude power-law fits.
    For $A,B$ each one site the mana density is $m_{\textrm{cc}}(A,B) \lesssim 10^{-14}$
    (numerically zero)
    for $\delta x > 60$,
    because $\rho_{AB}$ enters $\mathsc{STAB}$
    as it approaches $\rho_{A} \otimes \rho_B$.
    For $A,B$ each two sites and $\delta x \gg 1$ we find good agreement with the predicted exponent $\delta x^{-4/15}$,
    though we cannot rule out other small exponents.
  }
\label{fig:twopoint}
\end{figure}

Can the magic we saw in Sec.~\ref{ss:results:simply-connected} be removed by local unitaries?
If it can, it is of limited relevance to state preparation,
because we can choose to perform our quantum computation in a low-magic basis;
and to AdS-CFT, because the magic will be a lattice-scale, non-universal effect.

To probe the degree to which the mana is nonlocal
(that is, whether it can be removed by local unitaries)
we measure the ``connected component of mana density''
\begin{align}
  \label{eq:mcc-def}
  m_{\mathrm{cc} }(A,B) = m(\rho_{A\cup B}) - \frac 1 2 [m(\rho_{A}) + m(\rho_{B})]
\end{align}
on two widely-separated subregions $A$ and $B$.\footnote{
  This is a heuristic diagnostic of ``nonlocal magic''.
  We plan to address the degree to which it reflects how much of the system's magic can be removed by local unitaries---and the degree to which it is a physically meaningful notion of ``nonlocal magic''---in future work.
}
The factor of $1/2$ comes about because we measure mana density, not mana.
We do this for ease of comparison between subregion sizes, and with Sec.~\ref{ss:results:simply-connected}.

Fig.~\ref{fig:twopoint} shows the connected component of mana density
for the case where $A, B$ are 1-site subregions (top)
and for the case where $A, B$ are subregions with 2 adjacent sites (bottom)
as a function of the separation $\delta x$ between the two sites $i$ and $j$.
(In the 1-site subregion case $m(\rho_A) = m(\rho_B) = 0$; in the two-site subregion case, we approximate $m(\rho_B) \simeq m(\rho_A)$ for convenience.)

In each case the connected component of mana density shows a clean separation between $\theta < \pi/4$, $\theta = \pi/4$, and $\theta > \pi/4$. 
For $\theta \ll \pi/4$ the mana drops rapidly to zero,
because for $\delta x \gtrsim \xi$ the localization length,
$\rho_{AB} \simeq \rho_A \otimes \rho_B$ and $\mathcal M(\rho_{AB}) \simeq \mathcal M(\rho_A) + \mathcal M(\rho_B)$.
Conversely for $\theta \gg \pi/4$ the mana density $m(\rho_{ij})$ plateaus,
because the state is long-range entangled.

To understand Fig.~\ref{fig:twopoint} in some detail,
we construct a toy model for the reduced density matrix $\rho_{A\cup B}$.
(We write the toy model on two sites,
but extend in the natural way to four sites for $A,B$ are each two sites.)
For $\theta = \pi/2$ it will be
\begin{equation}
  \ket{\Omega_0} = \sum_n X_i^n X_j^n\ket{00}\;.
\end{equation}
But for $\theta < \pi/2$,
the $X_j$ term will cause the $Z$ eigenstates to hybridize with each other,
and cause the two-site subsystem to be more mixed.
Mock this up by writing
\begin{equation}
  \label{eq:toy}
  \rho_{\toy} = \frac 1 3 (1 - \alpha_{ij}) \rho_i\otimes \rho_j + \alpha_{ij} \ketbra{\mathbb N_2}{\mathbb N_2}
\end{equation}
with
\begin{equation}
 \ket{\mathbb N_2} \propto \sum_n X^n (2\ket {00} - \ket {11} - \ket {22})\;.
\end{equation}
(up to normalization).
The connected correlation function is
\begin{equation}
  \expct{Z^\dagger_i Z_j + h.c.} - 2\Re[\langle Z^\dagger_i\rangle\expct{Z_j}] \propto \alpha_{ij}\;,
\end{equation}
so take
\begin{align}
  \alpha \propto
  \begin{cases}
    b + c e^{- |j - i|/\xi} & \theta > \pi / 4\\
    |j - i|^{-2\Delta} & \theta = \pi/4\\
    e^{-|j-i|/\xi} & \theta < \pi/4
  \end{cases}
\end{align}
with $2\Delta = 4/15$ \cite{wu_potts_1982}.

For $A$ and $B$ each two sites, this is the whole story.
The connected component of mana measures how far $\rho_{A \cup B}$ is from $\rho_{A} \otimes \rho_B$, and we predict
\begin{equation}
  m_{\mathrm{cc}}(A) = C_1 \delta  x^{-2\Delta} - \varepsilon\;,
\end{equation}
where we allow ourselves an offset $\varepsilon$ to account for our approximation $m(\rho_B) \approx m(\rho_A)$.
This is consistent with we see in Fig.~\ref{fig:twopoint},
though we emphasize that we would need much larger system sizes and much more precise DMRG calculations to rule out other exponents.

For $A$ and $B$ each one site the situation is more complicated.
The one-site reduced density matrices have
\begin{equation}
  m(A) = m(B) = 0;
\end{equation}
we take them to be in $\mathsc{stab}$.
The convex hull $\mathsc{stab}$ is a polytope.
As $\alpha$ decreases, $\rho_\toy(\alpha)$ passes through a facet of $\mathsc{stab}$ at some finite $\alpha_{0}$.
Immediately before $\alpha_0$, then, the distance to the nearest stabilizer state is
\begin{align}
  \label{eq:toy-wigner-dist}
  \begin{split}
    d_S &:= \min_{\sigma\in \mathsc{stab} } \| \rho_\toy(\alpha) - \sigma\|\\
    &\sim (\alpha  - \alpha_{0})\\
    &\sim C_1 \delta x^{-2\Delta} - C_2\;.
  \end{split}
\end{align}
But the mana is more complicated.
Up to a Hilbert space dimension factor,
mana is the $L^1$ (taxicab) distance to the nearest operator with positive Wigner representation,
and the set of operators with positive Wigner representation is larger than the set of states with positive Wigner representation,
which in turn is slightly larger than $\mathsc{stab}$.
For $\delta x$ not too large
the state is far from all of these sets,
and the distinction does not matter.
But we expect the behavior of Eq.~\eqref{eq:toy-wigner-dist} only for $\delta x \gg 1$,
when lattice-scale effects are unimportant.
This suggests that the details of the geometry of the space of Wigner functions
determine the value of $m_{\mathrm{cc}}(A,  B)$ for $A,B$ each one site;
the only concrete prediction we can make is that it will be identically zero at some finite separation $\delta x$.

This is exactly what we see in Fig.~\ref{fig:twopoint} (top):
the mana is numerically zero for $\delta x \ge 60$.
Empirically, our data is consistent with 
\begin{equation}
  m_{\mathrm{cc}}(A,B) = C_1 \delta x^{\mu} - C_2\;,\quad \mu = \frac 1 {20}\;,
\end{equation}
though (again) we do not claim to have precisely determined the exponent,
nor do we claim that a power-law ansatz captures the interesting details of this quantity's behavior.

The disappearance of the mana for finite $\delta x = |i - j|$ in the one-site case is analogous to the ``sudden death of entanglement'' of \cite{yu_sudden_2006}, so we call it ``sudden death of magic''.
In sudden death of entanglement, a stochastic magnetic field moves two entangled qubits into the region of zero Wootters concurrence (hence zero entanglement) in finite time.
In Sec.~\ref{ss:mera:twopoint}, we will see that the two-site reduced density matrix of the Potts ground state is the result of repeated application of a quantum channel derived from a MERA representation of that ground state;
this plays the role of the stochastic magnetic field in \cite{yu_sudden_2006}.

\section{Magic in MERAs}\label{s:mera}
\begin{figure}
  \includegraphics[width=0.45\textwidth]{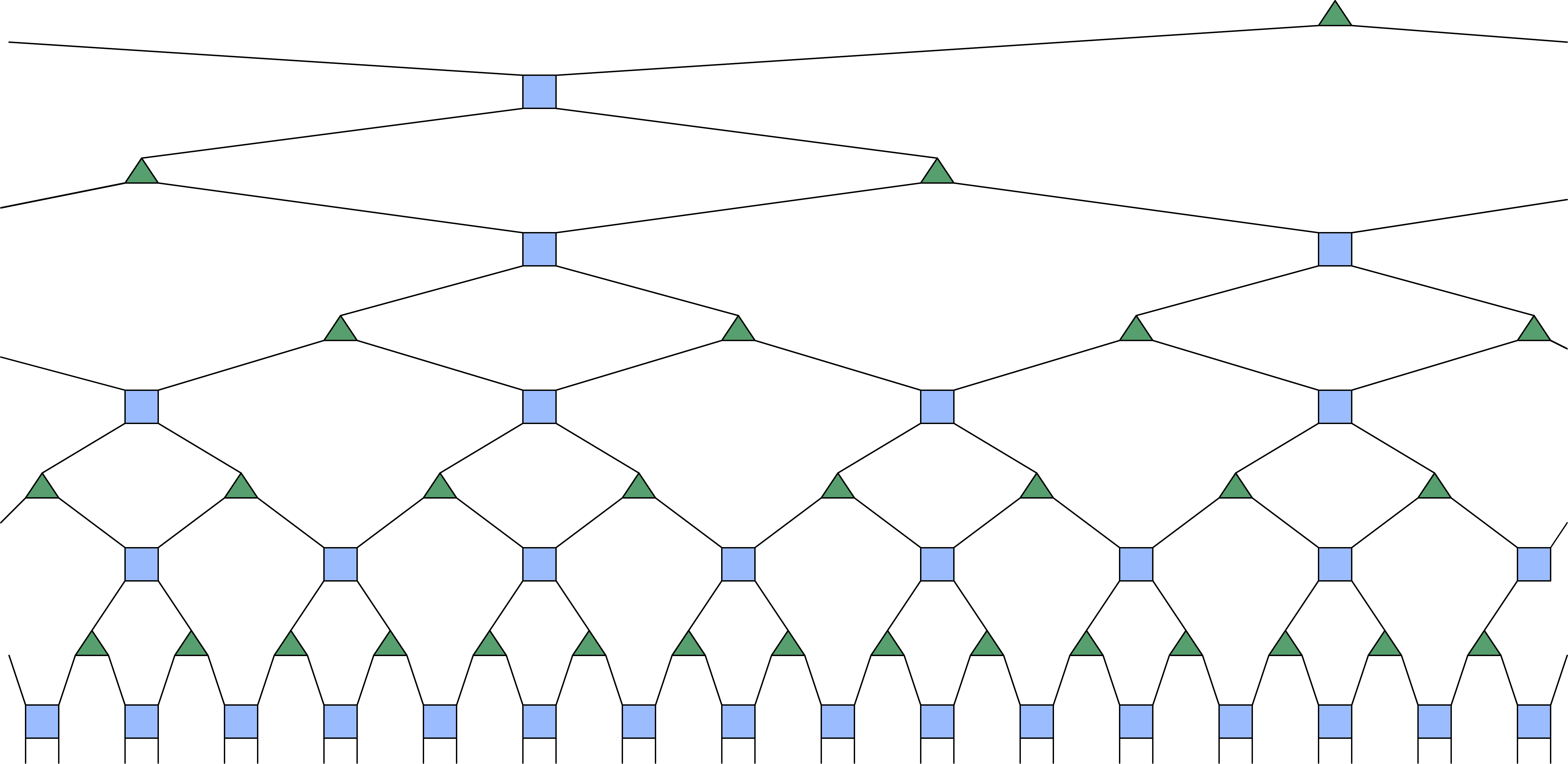}
  \caption{\textbf{MERA representation} of a critical ground state.
    In this work, we read a MERA from top to bottom as a quantum circuits preparing the state in question.
    The blue squares are unitary ``disentanglers'';
    the green triangles are ``isometries'' that
    embed a small Hilbert space in a larger Hilbert space. The isometries can always be re-written as unitary gates by expanding the (smaller) input Hilbert space. This is guaranteed by the Stinespring dilation theorem.
    The network structure results in
    (1) scale invariance,
    (2) entanglement logarithmic in subsystem size,
    (3) power-law correlation functions, and
    (4) concrete predictions for subsystem and two-point mana.
  }
  \label{fig:whole-mera}
\end{figure}

\begin{figure*}[t]
  \begin{minipage}{0.45\textwidth}
    \includegraphics[width=\textwidth]{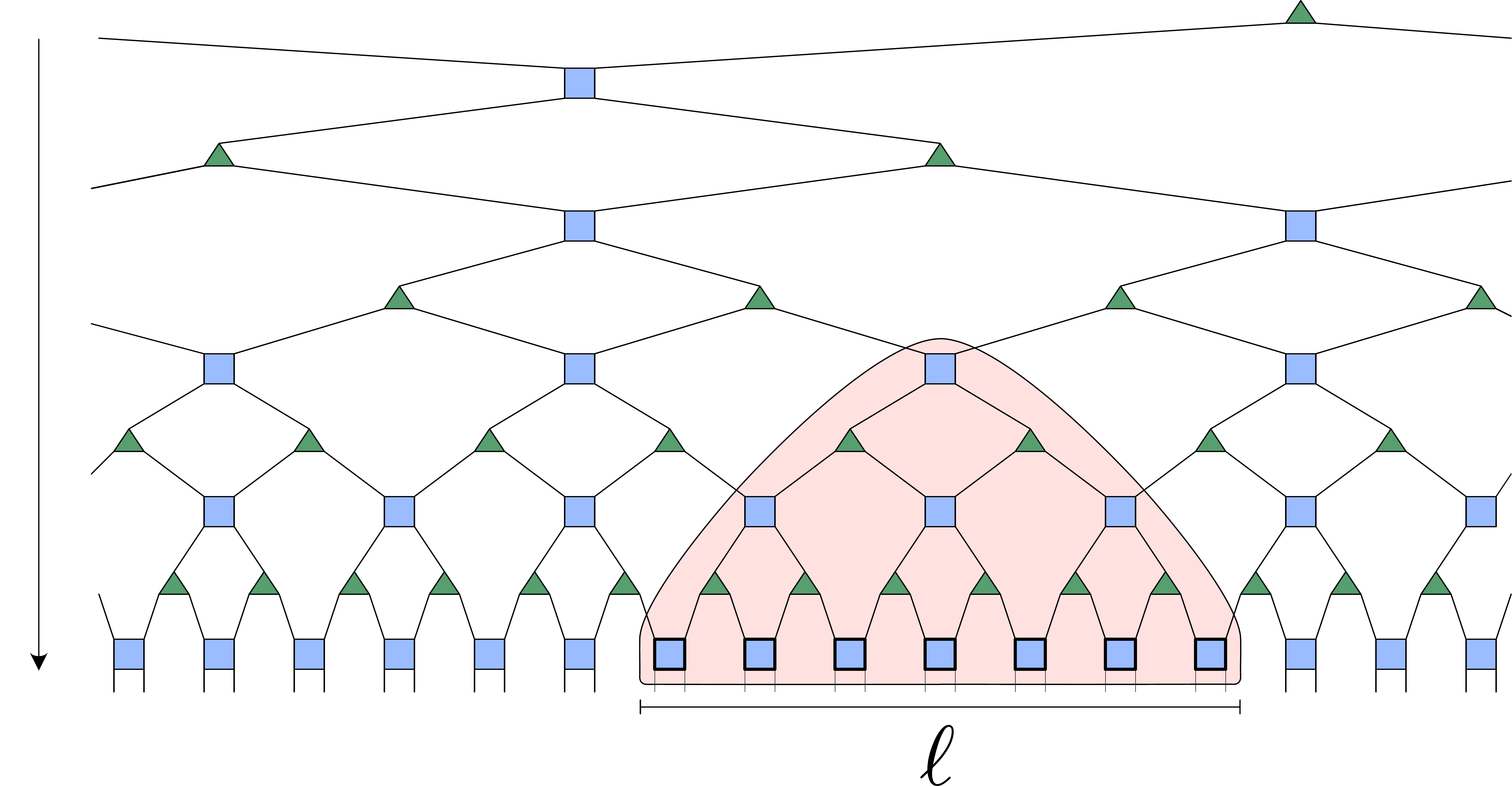}
  \end{minipage}
  \begin{minipage}{0.45\textwidth}
    \includegraphics[width=\textwidth]{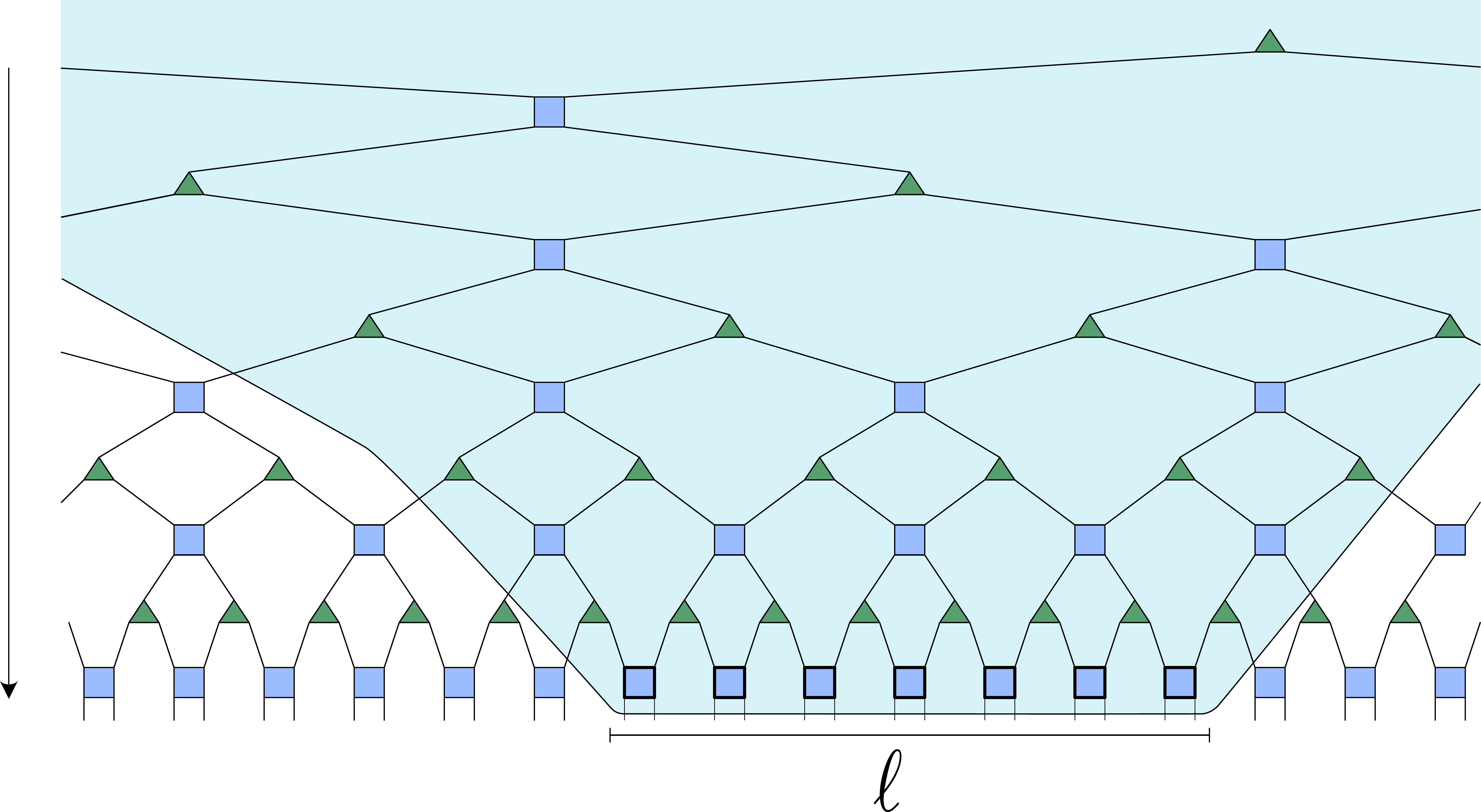}
  \end{minipage}
  \caption{\textbf{Causal domains of an $l$-site subsystem} of a MERA state.
    \textbf{Left},
    the ``past domain of dependence'':
    the set of gates that influence only the subsystem in question,
    and no other degrees of freedom.
    \textbf{Right},
    the ``past causal cone'':
    the set of all gates that influence the subsystem in question,
    whether or not they also influence other gates.
  }
  \label{fig:mera:causal-domains}
\end{figure*}

The results and toy model of Sec.~\ref{s:results}
appear naturally in
a picture of the CFT ground state as a MERA.
In this section we estimate the magic by understanding the MERA as a quantum circuit that prepares the CFT ground state
and counting non-Clifford rotations.

Because we imagine counting non-Clifford rotations in a circuit,
these estimates are (roughly) upper bounds on the number of non-Clifford rotations required to prepare the state.
In principle, clever recompilation could reduce that number.
But the numerical measurements of mana in Sec.~\ref{s:results} are (heuristically) lower bounds on the number of non-Clifford rotations required to prepare the state (recall Eq.~\eqref{eq:mana-bound} in Sec.~\ref{s:mana-bound}).
So the fact that the MERA estimates of this section broadly match those numerical measurements of Sec.\ref{s:results}
indicates that the MERA circuits are close to optimal,
and much of the magic is best understood as nonlocal---as residing in gates that affect many scales, including the largest.

We show the MERA for a CFT ground state in Fig.~\ref{fig:whole-mera}.
The blue squares are ``disentanglers'',
while the green triangles are ``isometries'';
the disentanglers and isometries satisfy
\begin{align}
  \begin{split}
    \diag{7ex}{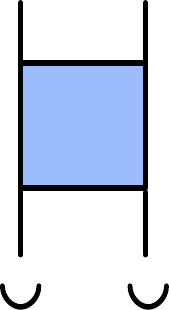} = \diag{0.5ex}{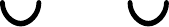}\\
    \diag{7ex}{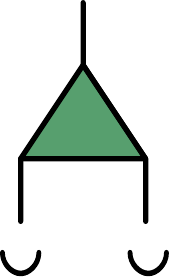} = \diag{0.5ex}{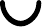}\;.
  \end{split}
\end{align}
We use the notation $\diag{0.5ex}{isometry-cap-R}$ for traces over the relevant legs.
The result is (implicitly) a sandwich structure like
\begin{equation}
  \diag{7ex}{isometry-cap-L} \quad \equiv \quad \diag{4ex}{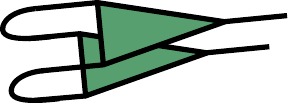}\;.
\end{equation}

We can identify a causal structure on the MERA by re-interpreting it as a quantum circuit in either of two ways.
If we read it from bottom to top, the MERA implements a renormalization procedure:
each twofold layer,
comprising a set of unitaries and a set of isometries,
projects out UV degrees of freedom, leaving only IR information.
If we read it from top to bottom, the MERA offers
a recursive, scale- and translation-invariant recipe for constructing the CFT ground state:
each twofold layer doubles the number of degrees of freedom (implicitly by taking the tensor product of each site with $\ket 0$),
and fills in the details in a way that is
(1) consistent with the IR information coming from the layer above, and
(2) scale-invariant. 
We use the latter ``top-to-bottom'' or ``recipe'' reading.

In Sec.~\ref{ss:mera:finite},
we use the causal structure
imposed by the ``top-to-bottom'' reading of the MERA as a quantum circuit
to estimate the mana of finite subsystems.
The resulting form matches the results of Sec.~\ref{ss:results:simply-connected}.

In Sec.~\ref{ss:mera:twopoint} we then sketch a calculation of the mana of reduced density matrices on widely separated regions.
That calculation has a straightforward structure
(repeated application of a quantum channel);
this structure---together with some input from the numerics of Sec.~\ref{s:results}
about simply connected subsystems---
strongly constrains the MERA prediction of the connected component of magic.
The resulting form matches the results of Sec.~\ref{ss:results:twopoint}

Finally, in Sec.~\ref{ss:mera:nearcritical},
we use quasi-MERA structures for ground states near but not at the critical point
to estimate the mana in that regime.

We have not constructed this MERA for a particular CFT (though we look forward to future work doing so),
so we cannot explicitly evaluate the tensor network,
but we can argue based on its structure and generic assumptions about the properties of MERAs.
We therefore expect these observations to be applicable for all CFTs that can be efficiently approximated by MERAs.

\subsection{Finite subsystems}\label{ss:mera:finite}

\begin{figure*}
  \includegraphics[width=\textwidth]{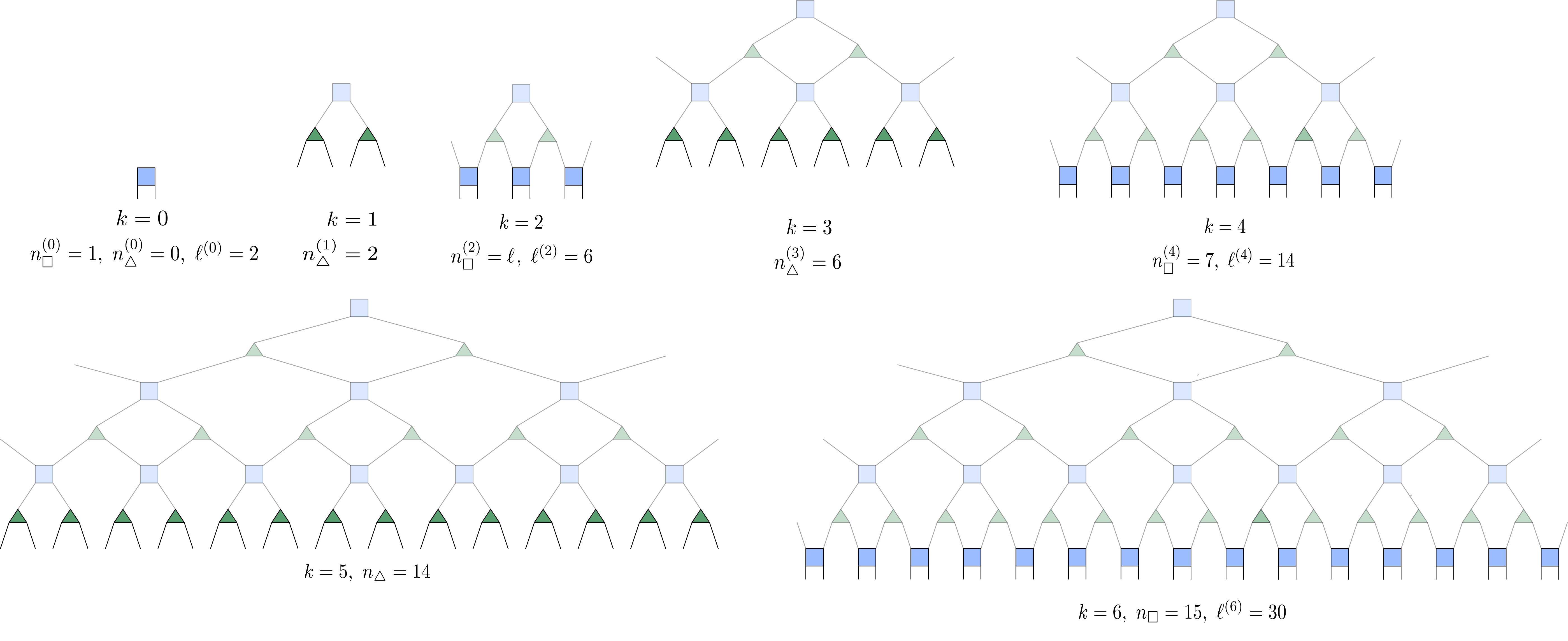}
  \caption{\textbf{Iteratively constructing larger regions with larger past domains of dependence}
    (cf Fig.~\ref{fig:mera:causal-domains} left, Sec.~\ref{ss:mera:finite}).
    Interpret the MERA as a top-to-bottom specification for a circuit that---after each set of blue square disentanglers---produces a valid ground state for the CFT.
    Start with a single blue disentangler; this is layer $k = 0$,
      and the past domain of dependence of two sites.
    Add two green isometries and then three blue disentanglers; this is layer $k = 2$,
      and the past domain of dependence of two sites.
    Proceeding in this way, at layer $2k-1$ we add $n_\triangle^{(2k-1)} = 2^{k+1}-2$ green isometries,
      and at layer $2k$ we add $n_\square^{(2k)} = 2^{k+1} - 1$ blue disentanglers,
      at which point we have constructed the past domain of dependence for $\ell^{(2k)} = 2^{2k+2} - 2$ sites.
  }
  \label{mera:past-domain-construction}
\end{figure*}

In this section we estimate the mana of simply-connected subsystems of a MERA by counting gates,
and compare to the mana of finite subsystems of the MPS ground states of Sec.~\ref{s:results}.
This is a not a stringent test of the MERA ansatz!
We have only $\ell = 1 \dots 7$-site subsystems.
Many sensible finite-size scaling predictions
can be made to match for such small system sizes.
Moreover our estimate for the MERA is very heuristic.
Nonetheless, it is important to check that the MERA is at least consistent in this way with our numerics.

A state's mana estimates the number of T gates
(or, more generally, non-Clifford rotations)
required to produce that state.
Given a circuit that produces the state,
we can estimate the magic
by counting the number of non-Clifford rotations
in the circuit. 
The MERA Fig.~\ref{fig:whole-mera} is just such a circuit.
If each disentangler or isometry introduces a certain amount of mana $\mu$,
the mana is roughly
\begin{equation}
  \mathcal M \sim \mu \times [\text{\# of tensors}]\;.
\end{equation}

This is a deceptively simple prescription:
which gates, exactly, contribute to the mana of a finite subsystem of a MERA state?
Two options present themselves: the ``past domain of dependence'' and the ``past causal cone''.
The past domain of dependence is the set of gates
that influence only the subsystem in question,
and no others (Fig.~\ref{fig:mera:causal-domains} left),
while the past causal cone is the set of all tensors that influence the region in question (Fig.~\ref{fig:mera:causal-domains} right).
The number of tensors in the past causal cone diverges (logarithmically) with the size of the whole system;
since we expect the mana density of a finite subsystem to be independent of system size (and indeed the mana density is bounded by $\frac 1 2 \log 3$),
this is unsuitable.
We therefore use the past domain of dependence. Intuitively, this is also a sensible choice because tensors that are inside the past causal cone but outside the past domain of dependence influence many different subsystems, in addition to the subsystem of interest.
So the magic contribution from such tensors should be ``diluted'' among these systems, and thus only subleading to the past domain of dependence contribution.
(Additionally, the form of the result turns out not to be sensitive to small changes in the domain in which we count.
We discuss this further below.)

To estimate the number of tensors in certain regions' domains of dependence
proceed inductively, as in Fig.~\ref{mera:past-domain-construction}.
Instead of specifying larger and larger subsystems and finding their past domains of dependence,
iteratively construct larger and larger past domains of dependence and measure the size of the regions whose domains they are.
Start with two sites whose domain of dependence consists of a single blue square disentangler.
This domain of dependence has
\begin{equation}
  n^{(0)}_\square = 1\ \text{disentanglers}\;.
\end{equation}
Add a layer (layer 1) of $n^{(1)}_\triangle = 2$ isometries and a layer (layer 2) of $n^{(2)}_\square = n^{(1)}_\triangle + 1 = 3$ disentanglers;
this domain of dependence corresponds to a region of size $\ell^{(2)} = 2n^{(2)}_\square = 6$ sites.
Proceeding in this way,
each disentangler in layer $k$ gives rise to
\begin{equation}
  \label{eq:rec-tr}
  n_\triangle^{(k+1)} = 2n^{(k)}_\square
\end{equation}
isometries in layer $k+1$,  which in turn give rise to
\begin{equation}
  \label{eq:rec-sq}
  n_\square^{(k+2)} = n^{(k+1)}_\triangle + 1
\end{equation}
disentanglers; the region corresponding to this domain of dependence
has $\ell^{(k+2)} = 2n^{(k+2)}_\square$ sites.
Solving the recursion relation \eqref{eq:rec-tr},\eqref{eq:rec-sq}, we find
\begin{align}
  \begin{split}
    n^{(2k-1)}_\triangle &= 2^{k+1} - 2\\
    n^{(2k)}_\square &= 2^{k + 1} - 1\\
    \ell^{(2k)} &= 2\cdot(2^{k+1} - 1)\;.
  \end{split}
\end{align}
If disentanglers and isometries each contribute mana $m_\square, m_\triangle$,
then the mana of an $\ell^{(2k)}$-site subsystem should be roughly
\begin{align}
  \begin{split}
    M^{(2k)} &= m_\square n^{(0)}_\square +  \sum_{k' = 1}^k [m_\triangle n_\triangle^{(2k-1)} + m_\square n^{(2k)}_\square] \\
    &= m_\square - k (m_\square + 2 m_\triangle) + 4 (m_\triangle + m_\square) (2^k - 1)
  \end{split}
\end{align}
or (dropping the ${}^{(2k)}$ label) a mana density
\begin{align}
  \begin{split}
    \label{eq:mera:finite-mana-density}
    m(\ell) &= m_\square + m_\triangle\\
    &\qquad- \ell^{-1}(m_\square + 2m_\triangle) \big[\lg[(\ell + 2)/4] + 1\big]
    \end{split}
\end{align}

\begin{figure}
  \includegraphics[width=0.45\textwidth]{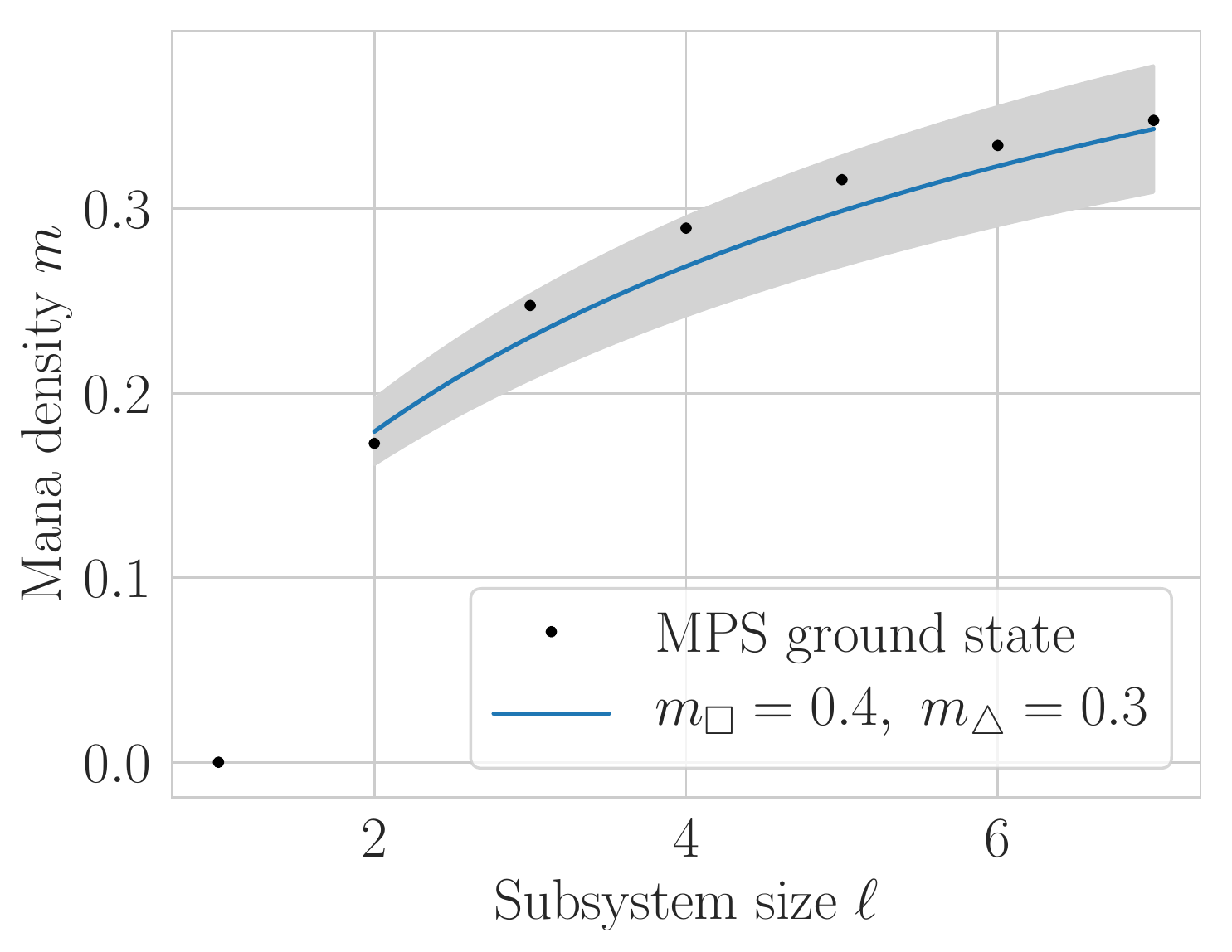}
  \caption{\textbf{Subsystem mana density as a function of subsystem size} at $\theta = \pi / 4$.
    The black dots show MPS measurements.
    The blue line shows the MERA estimate for $m_\square = 0.4, m_\triangle = 0.3$;
    the shaded region shows a $10\%$ variation in $m_\square, m_\triangle$,
    to emphasize the heuristic nature of our fit.
  }
  \label{fig:mera:data:finite}
\end{figure}
We plot the resulting prediction
against the $\theta = \pi/2$ finite-subsystem data of Sec.~\ref{s:results}
in Fig.~\ref{fig:mera:data:finite}.
We take $m_\square = 0.4$ and $m_\triangle = 0.3$, fit by eye, but we plot a range of $m_{\square,\triangle}$
to emphasize the rough nature of our fit.

To understand the effect of our choice to count tensors in the past causal cone,
imagine perturbing the boundary outward by an $O(1)$ number of tensors connecting to each leg in the domain of dependence.
The resulting ``boundary effect'' will increase the magic by
\begin{equation}
  \delta m_{\text{perturbation}}(\ell) \sim \ell^{-1} \ln \ell\;:
\end{equation}
that is, it will renormalize the coefficient of the finite size correction
without substantially changing its form. 

Two anomalies are apparent in Fig.~\ref{fig:mera:data:finite}.
First, the MPS measurements bend over faster than the MERA prediction.
Second, asymptotic value of the MERA prediction $m_\square + m_\triangle = 0.7$
breaks the Jensen's inequality bound $m \le \frac 1 2 \ln 3 \simeq 0.55$.
Both these anomalies result from the fact that repeatedly applying magic gates
does not always increase the system's mana.
(This is essentially the statement that a circuit with many T gates
can sometimes be recompiled to reduce the number of T gates,
but one can see it concretely in the case of Haar-random unitaries:
applying one Haar-random unitary to a computational basis state
will likely increase the mana to near its maximal value,
but applying a second Haar-random cannot increase it much more. )
Consequently finite subsystems contain more of the whole-system mana density
than one might expect from our MERA picture.
We conjecture that the mana density of the critical state is in fact maximal.

\subsection{Two-point magic}\label{ss:mera:twopoint}

\begin{figure}[t]
  \includegraphics[width=0.45\textwidth]{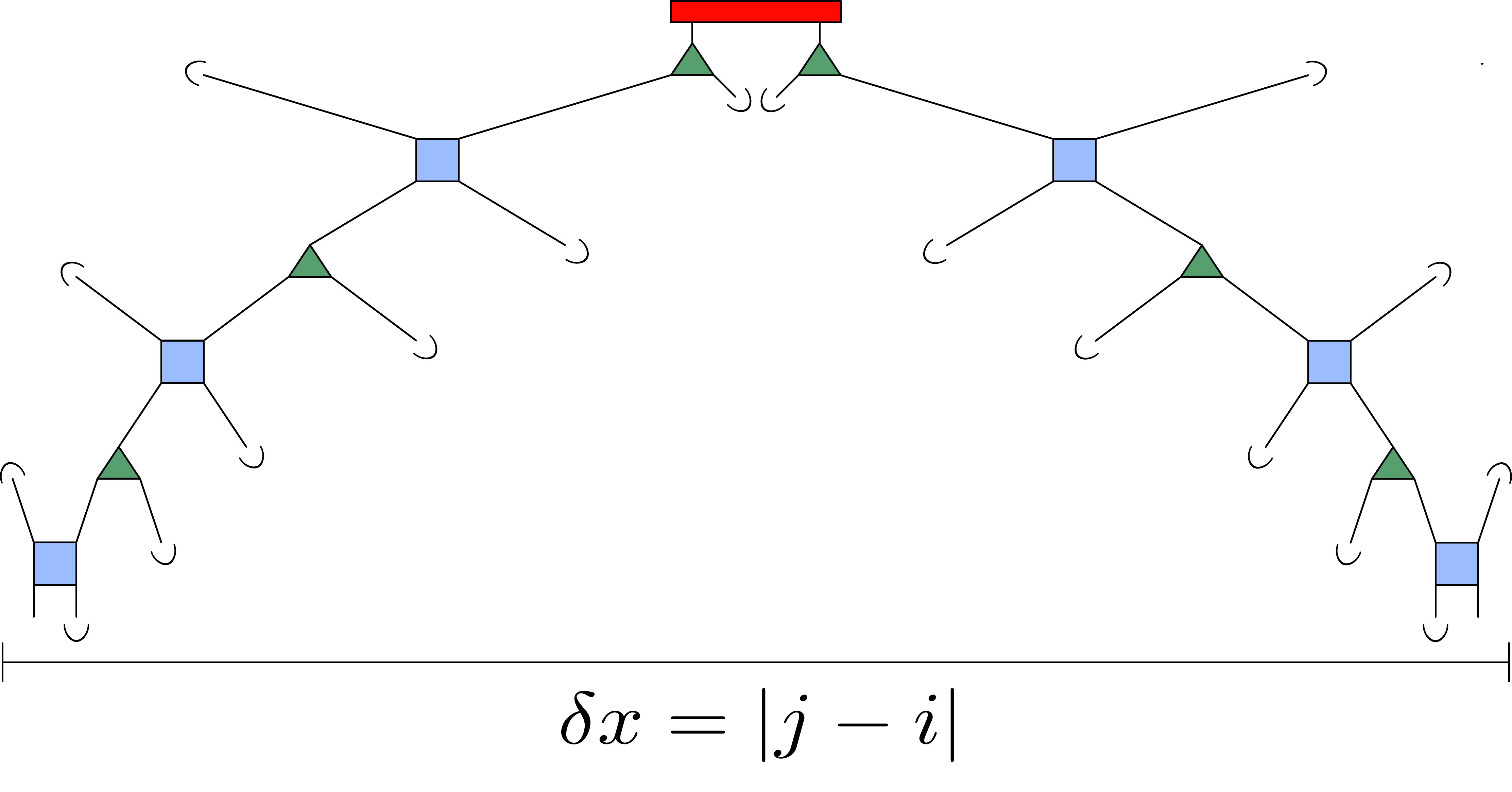}
  \caption{
    \textbf{Two-site reduced density matrix}
    resulting from tracing out physical degrees of freedom other than sites $i,j$.
    in the MERA of Fig.~\ref{fig:whole-mera}.
    Tracing out sites in $i < p < j$ in between $i$ and $j$ is equivalent to tracing out the indicated downward-facing legs, using the fact that the intervening tensors are unitaries or isometries.
    The effect of tensors outside the the past domain of dependence of the region $[i,j]$ is more subtle.
    We argue that they have little effect on correlations,
    and we imagine accounting for their effect on local properties
    by using empirical simply-connected-region density matrices taken from the MPS calculations of Sec.~\ref{s:results}.
    Even with the full two-site reduced density matrix, we can replace the above simplified channel with the actual descending super-operator~(see \cite{Vidal2007}), for which the same arguments below hold. 
  }
  \label{fig:twosite-mera}
\end{figure}

We can make a somewhat more careful estimate of the two-point magic, that is
\begin{equation*}
  m\left(\tr_{\overline{A\cup B} } \ketbra{\Omega}{\Omega}\right)\;,
\end{equation*}
for widely-separated subsystems $A,B$,
by estimating that reduced density matrix.
Fig.~\ref{fig:twosite-mera} shows the key structure.
For simplicity we will take $A$ and $B$ each to be single-site ``subregions'' $i$ and $j$.
To extend our argument to multi-site subregions, group nearby sites.

The reduced density matrix on those two sites,
separated by a distance $\delta x = |j - i|$,
is given by contracting the tensors that are outside the two-site past causal cone in the MERA.
Although the two sites in the reduced density matrix do not share a non-trivial past domain of dependence when they are not adjacent to each other,
it is possible to find a coarse-grained layer $\sim \log\delta x$ layers above,
where their nearest ancestors are adjacent.
Let us call this layer the ``parent layer'' and the reduced state of the adjacent ancestors the ``parent reduced state''.\footnote{This ``parent reduced state'' will in fact be like the model's ground state on a small system, but its nature is not important.}
The latter is shown in red in Fig.~\ref{fig:twosite-mera}.
From Sec.~\ref{ss:mera:finite},
we know the parent reduced state has non-zero amount of magic mostly coming from the tensors in its past domain of dependence.
Tracing over the in-between sites $i < p < j$,
we can construct a circuit (Fig.~\ref{fig:twosite-mera}) that relates the magic in the reduced density matrix $\rho(\delta x)$ to the parent reduced state. 


To be more explicit: start with the past domain of dependence of the simply-connected subsystem $[i,j]$---the red region in Fig.~\ref{fig:mera:causal-domains},
now with $\ell := |j - i|$.
In Sec.~\ref{ss:mera:finite} we argued that the mana of this subsystem was determined by the tensors in the past domain of dependence.
To make this more precise,
imagine tensoring the ``parent reduced state''
$\diag{2ex}{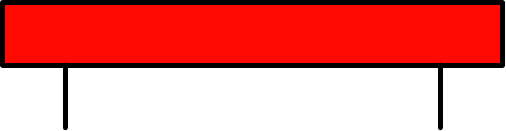}$
onto the top disentangler, which $i$ and $j$ have in common.
Then trace over the ``external legs'': the remaining legs connecting the domain of dependence to the rest of the MERA.
The tensors we ignore in this way will affect local properties but not---at leading order---correlations.
We further justify this construction and how to phenomenologically modify it below,
after we discuss its consequences.

With the external legs traced over,
the rest of the construction is straightforward:
trace out the physical sites $i < p < j$
and use the fact that the tensors are unitaries or isometries
to eliminate them.
The result is Fig.~\ref{fig:twosite-mera}.


Examining this construction, we see that the reduced density matrix
$\rho(\delta x) := \tr_{\overline{\{i,j\}}} \ketbra{\Omega}{\Omega}$
is the image of the parent reduced state $\diag{2ex}{twosite-mera-start}$
under $\log_2 \delta x$ iterations of the quantum channel
\begin{align}
  \label{eq:channel}
  \begin{split}
    D[\diag{1.5ex}{twosite-mera-start}] &\equiv (D_L \otimes D_R)[\diag{1.5ex}{twosite-mera-start}]\\
    &\\
    &=\ \diag{14ex}{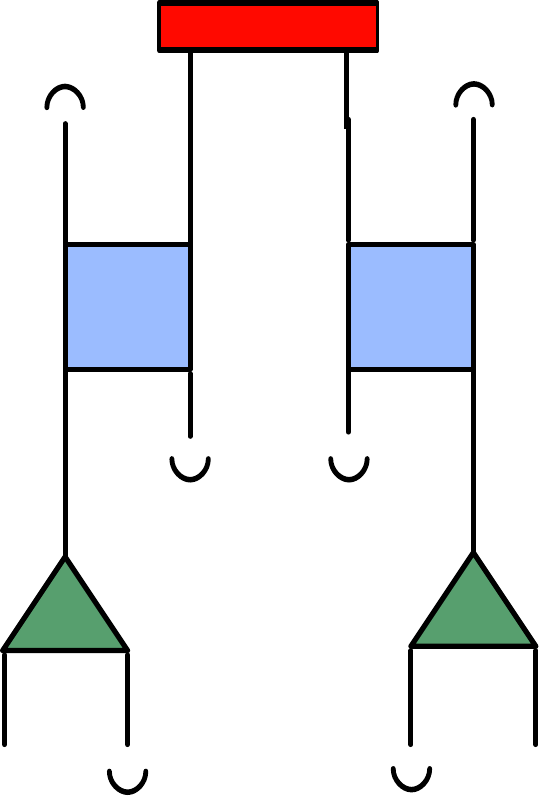}\;.
  \end{split}
\end{align}
We do not know this quantum channel explicitly,
because we do not know the MERA tensors.

\begin{figure}[t]
  \includegraphics[width=0.45\textwidth]{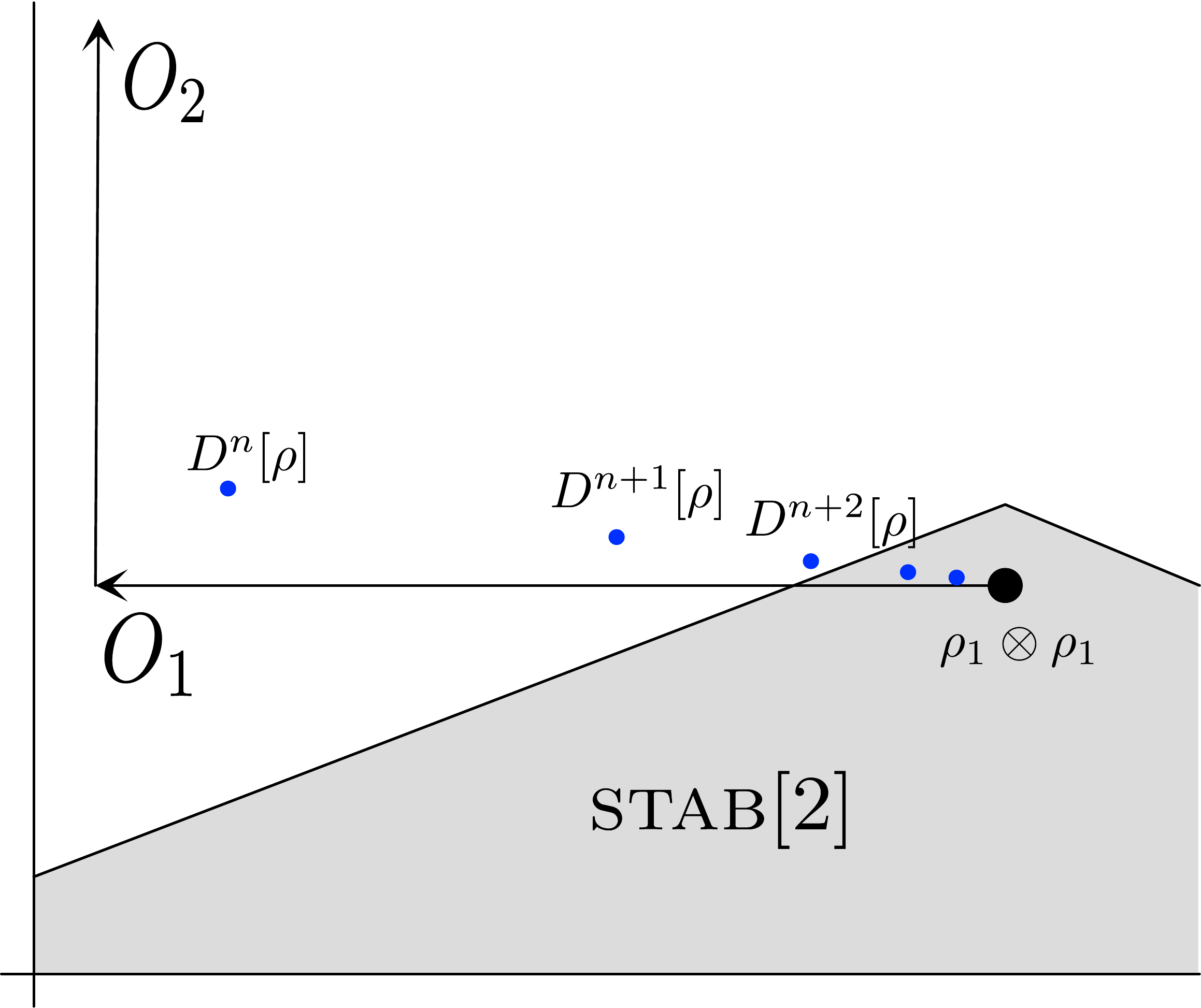}
  \caption{\textbf{Image of density matrices under iteration of the MERA channel}
    $D$, Eq.~\eqref{eq:channel} (cf Fig.~\ref{fig:twosite-mera})
    in the plane
    (two-dimensional affine space)
    defined by $\rho_1^{\otimes 2}, \rho_1^{\otimes 2} + O_1$, and $\rho^{\otimes 2} + O_1 + O_2$.
    The gray region is $\stab[2]$.
    The black dot shows $\rho_1^{\otimes 2}$,
    while the blue dots show $D^{n}[\rho] \dots D^{n + 4}[\rho]$.
    $D^{k}[\rho]$ asymptotically approachs $\rho_1^{\otimes 2}$ (cf Eq.~\ref{eq:approach}),
    but---because $\rho_1^{\otimes 2}$ is inside $\stab[2]$---$D^{k}[\rho]$ enters $\stab[2]$ after a finite number of steps $k$.
  }
  \label{fig:twosite-approach}
\end{figure}

 But we do know that its fixed point is two copies of the one-site reduced density matrix of the infinite-system ground state, which is close to the MPS of Sec.~\ref{s:results}.
Call that one-site density matrix $\rho_1$.
The channel therefore has spectrum
\begin{equation}
  \begin{split}
    D[\rho_1\otimes\rho_1] = \rho_1\otimes\rho_1\\
    D[O_j] = \lambda_j O_j
  \end{split}
\end{equation}
for some eigenoperators $O_j$, $\tr O_j^\dagger O_k = \delta_{jk}$, and eigenvalues
\begin{equation}
  \lambda_{q^4-1} < \dots < \lambda_1 < 1\;.
\end{equation}
(The strict inequality $\lambda_1 < 1$ follows because the state does not have long-range correlations.)
So for separation $\delta x \gg 1$,
\begin{equation}
  \label{eq:approach}
  \rho(\delta x) \sim \rho_1 \otimes \rho_1 + C \lambda^{\lg \delta x} = \rho_1\otimes\rho_1 + C \delta x^{-2\Delta}\;.
\end{equation}
If $\rho_1 \otimes \rho_1$ is non-magical and in addition has a non-magical ball around it,
the distance between $\rho(\delta x)$ and the nearest point in $\mathsc{stab}[2]$ is
\begin{align}
  \label{eq:approach}
  \begin{split}
    \min_{\sigma \in \stab[n]} \|\rho(\delta x) - \sigma \|^2 &\sim C'^2 \lambda^{\frac 1 2 \lg \delta x} - B \\
    &\sim A\; \delta x^{-2\Delta} - B
    \end{split}
\end{align}
(cf Fig.~\ref{fig:twosite-approach}),
and all of our comments in Sec.~\ref{ss:results:twopoint} about the geometry of $\mathsc{stab}$ and of the set of operators with positive Wigner function apply.
But if $\rho_1 \otimes \rho_1$ has nonzero mana, then for $\delta x \gg 1$
the mana is
\begin{equation*}
  \mathcal M(\rho(\delta x)) - \mathcal M(\rho_1 \otimes \rho_1) \propto C \delta x^{-2\Delta}
\end{equation*}
by Taylor expansion.
Again, this is just as in Sec.~\ref{ss:results:twopoint}.

We can now understand how operators immediately outside the past domain of dependence of $[i,j]$ might modify our calculation.
They will change the local properties of the two sites:
that is, they will renormalize $\rho_1$ above.
But the leading-order contribution to the correlations comes from the shortest path through the lattice between the two sites.
This path is precisely that illustrated in Fig.~\ref{fig:twosite-mera},
whose effect is captured by the quantum channel $D$.
In this light our choice to take the eigenoperator $\rho_1$ of $D$ to be the empirical density matrix resulting from our matrix product state simulations takes on new significance.
$\rho_1$ captures all the local expectation values of the ground state.
So by taking it to be the fixed point of $D$,
we phenomenologically take into account the local effect of tensors outside the past domain of $[i,j]$.

\subsection{Mana near the critical point in a quasi-MERA}\label{ss:mera:nearcritical}

\begin{figure*}
  \includegraphics[width=\textwidth]{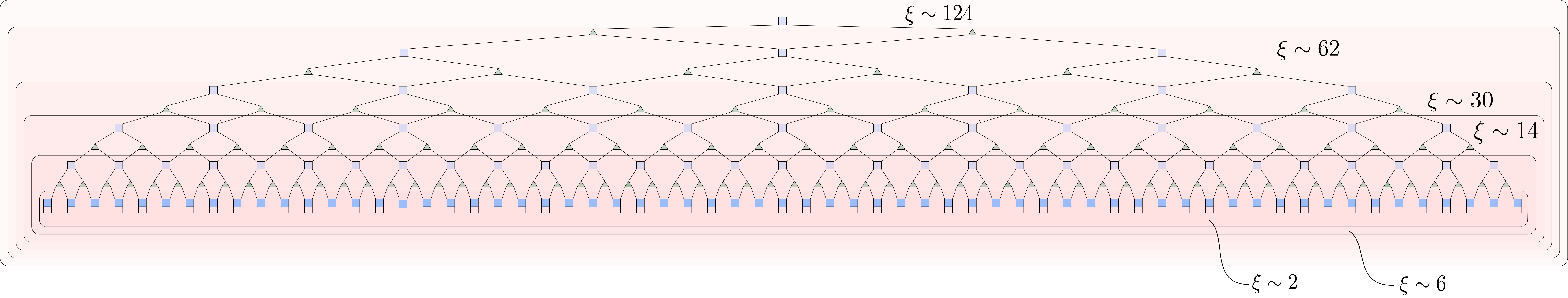}
  \caption{\textbf{Quasi-MERAs away from the critical point:}
    $|\theta - \theta_c|$ determines
    a correlation length $\xi \sim |\theta - \theta_c|^{-\nu}$,
    which in turn determines the number of MERA layers $k \sim \lg \xi \sim -\nu \lg |\theta - \theta_c|$
    required to approximate the state.
    We can therefore repeat the tensor-counting calculation of Sec.~\ref{ss:mera:finite}
    to estimate the mana near but not at the critical point.}
  \label{fig:mera:quasi}
\end{figure*}

Consider a length-$N$ system near but not at the critical point.
Were the system at the critical point,
it would be described by a MERA-like structure
with $2K+1$ layers, $\ell^{(2K)} = N$.
Since the system is not at the critical point
only the bottom $k$ of those layers remain,
where $k$ is controlled by the correlation length $\xi$ (cf Fig.~\ref{fig:mera:quasi}):
\begin{equation}
  \ell^{(2k)} \sim 2^k  \sim \xi \sim |\theta - \theta_c|^{-\nu} \;.
\end{equation}
We can repeat the analysis of Sec.~\ref{ss:mera:finite},
now starting the sum $k$ levels before the end:
the mana density of the infinite system is
\begin{align}
    m(k) &= \lim_{K \to \infty} \left([\ell^{(2K)}]^{-1}
      \sum_{k' = K - k}^K [m_\triangle n_\triangle^{(2k-1)} + m_\square n^{(2k)}_\square] \right)\notag\\
\end{align}
for
\begin{equation}
  m(\theta) \sim (m_\triangle + m_\square)( 1 - |\theta - \theta_c|^{\nu}) + \mathcal O(N^{-1})\;. \label{eq:qmera-pred}
\end{equation}
As we discussed in Sec.~\ref{ss:mera:finite},
beyond a certain point
(determined, roughly, by the bound $- \frac 1 2 \ln 3$ on the mana density)
additional MERA layers do not increase the magic.
So the prediction \eqref{eq:qmera-pred} should be modified to something like
\begin{equation}
  \label{eq:qmera-pred-mod}
  m(\theta)=\max\left[m_{\mathrm{max}}, (m_\triangle + m_\square)( 1 - |\theta - \theta_c|^{\nu})\right]\;,
\end{equation}
with $m_{\mathrm{max}} \sim \frac 1 2 \ln 3$;
the precise value of $m_{\mathrm{max}}$ could likely be estimated from a typicality calculation.

Eq.~\ref{eq:qmera-pred-mod} predicts the mana density of the whole system.
What does this picture predict for the mana density of a finite subsystem?
When the subsystem size is $\ell \gg \xi$
(that is, far from the critical point),
the subsystem's domain of dependence reaches all the way to the $k$-layer cutoff,
and the mana density is still given by Eq.~\ref{eq:qmera-pred}---that is, the finite subsystem effects of Sec.~\ref{ss:mera:finite} are not important.
When $\ell \ll \xi$, the subsystem's domain of dependence does \emph{not} reach the $k$-layer cutoff,
so finite-subsystem effects of Sec.~\ref{ss:mera:finite} dominate, and changing $\theta$ (hence $\xi$)
does not substantially change the finite-subsystem mana density.
The mana density will therefore be rounded at $\theta$ such that $\xi(\theta) \sim \ell$;
in Fig.~\ref{fig:contiguous} we shade the region $\xi(\theta) \gtrsim \ell$.
(This is essentially the usual story about finite-size scaling, rephrased in terms of MERAs and gates.)
In App.~\ref{app:corr} we find that  for
\begin{equation}
  |1 - \theta/\theta_c| \lesssim 7\times 10^{-2}
\end{equation}
the correlation length is $\xi \lesssim \frac 1 2 \ell_{\max} = 3.5$
where $\ell_{\max}$ is the size of the largest subsystem we treat,
so we expect good convergence in subsystem size.
This is broadly consistent with the behavior we see in Fig.~\ref{fig:contiguous}:
there we saw good convergence in subsystem size for $|1 - \theta/\theta_c|  > 7\times 10^{-2}$.

\section{Mean field study of the $q$ dependence of mana}\label{s:meanfield}

To complement our detailed study of mana in the $q=3$ case, it is interesting to consider the $q$ dependence of mana. The 1d Potts model has a first order phase transition for $q>3$. Because the transition is not associated with a diverging length scale, mean field theory should provide a reasonable qualitative guide to the physics. We will consider a general regular graph when setting up the mean field theory since the result depends only on the number of neighbors.

Let $G=(V,E)$ be a graph with degree $k$ for each vertex $v \in V$. A many-body Hilbert space can be constructed by assigning a $q$-dimensional Hilbert space to each vertex and taking the tensor product of these spaces. The Potts model on this graph is defined by the Hamiltonian
\begin{equation}
    H_G = - \sin \theta \sum_{(v,v')\in E} \left[Z_v^\dagger Z_{v'} + h.c.  \right]- \cos \theta \sum_{v\in V} \left[ X_v + X_v^\dagger \right].
\end{equation}
When $G$ is a linear chain, we recover the 1D Potts model.

The mean field ansatz is defined by a many-body state that is a tensor product of $|V|$ copies of the local state 
\begin{equation}
    \ket{\phi} = \alpha |Z=1\rangle + \sqrt{1-\alpha^2} \ket{\perp}\;,
\end{equation}
where the $\ket{\perp}$ is orthogonal to $\ket{Z=1}$ and given by
\begin{equation}
    \ket{\perp} = \frac{\sqrt{q}\ket{X=1} - \ket{Z = 1}}{\sqrt{q-1}}.
\end{equation}
The single variational parameter $\alpha$ turns out to be sufficient.

The mean field energy, which is the expectation value of $H_G$ in the state $|\phi\rangle^{\otimes |V|}$, is determined by the single site expectation values,
\begin{equation}
    \langle \phi | Z |\phi \rangle = \frac{q\alpha^2-1}{q-1}
\end{equation}
and
\begin{equation}
    \langle \phi | X |\phi \rangle = \frac{2 \alpha \sqrt{1-\alpha^2}}{\sqrt{q-1}} + (1-\alpha^2) \frac{q-2}{q-1}.
\end{equation}
The mean field energy per vertex is
\begin{equation}
    \frac{\langle H_G \rangle}{|V|} = -k \sin \theta \langle Z \rangle^2 - 2 \cos \theta \langle X \rangle.
  \end{equation}
  
  For each $\theta$, we optimize over $\alpha$
  with JuMP \cite{DunningHuchetteLubin2017,jump-devjumpjl_2020} and Ipopt \cite{wachter_implementation_2006,coin-oripopt_2020}
  to find the lowest energy state. The result is either paramagnetic, with $|\phi \rangle = |X=1\rangle$ and $\alpha = q^{-1/2}$, or ferromagnetic, with $\langle Z \rangle \neq 0$ and $\alpha >q^{-1/2}$. Numerically, for a given $q$, we find a transition at $\theta = \theta_c^{(q)}$. The transition is second order for $q=2$ and first order for $q>2$. Of course, this prediction is qualitatively incorrect when $G$ is a chain and $q=3$ as seen above, but the first order character for larger $q$ is qualitatively correct. 

\begin{figure}[t]
\includegraphics[width=0.45\textwidth]{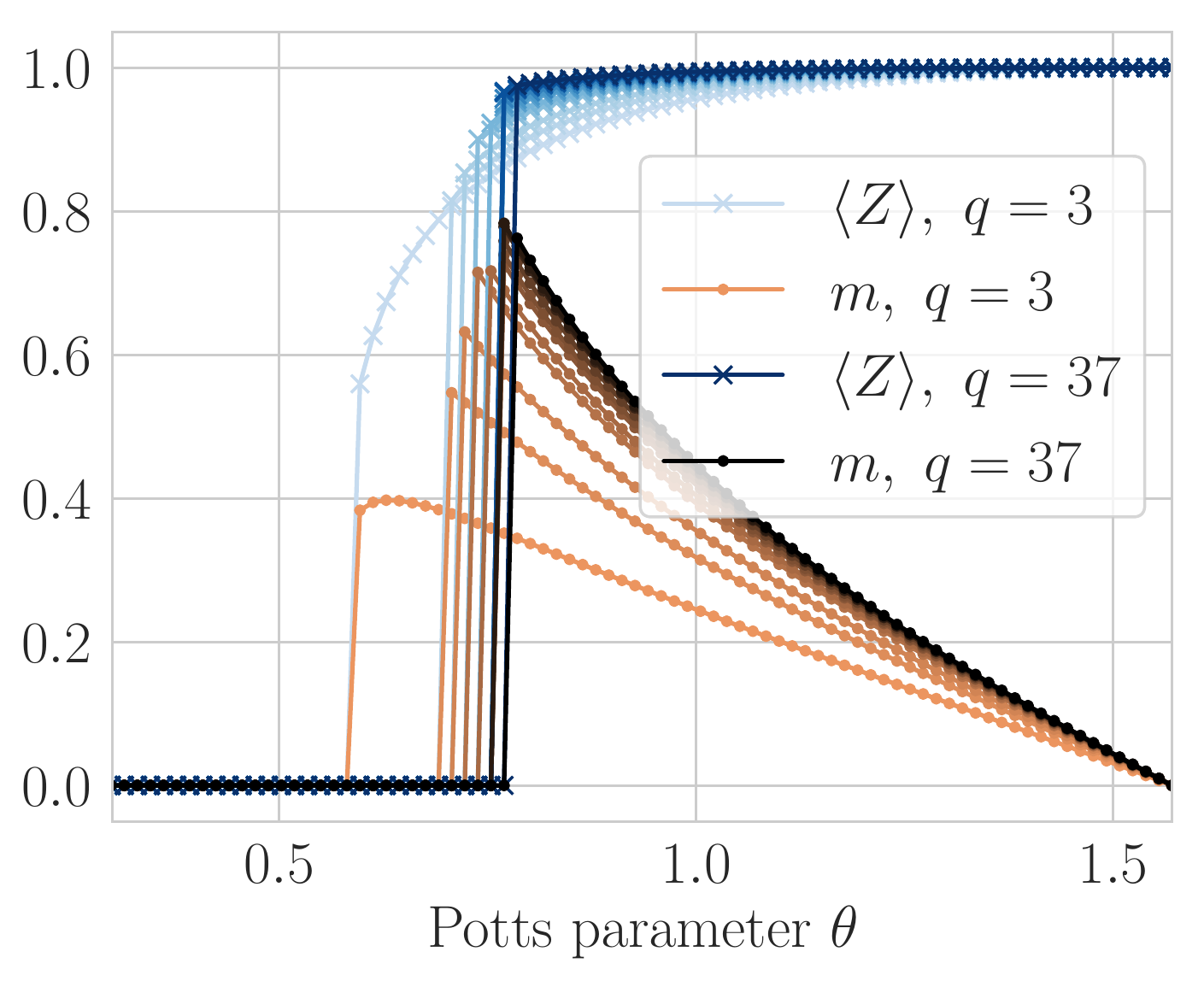}
  \caption{\textbf{Mana in a mean field theory} of the $q > 3$ Potts model.
    Mana $m(q;\theta)$ and magnetization $\expct{Z}$ as a function of $\theta$ for $q = 3,5,7,11,13,17, 19, 23, 29, 31, 37$. 
    The system displays a sharp jump between the paramagnetic and ferromagnetic phases;
    the peak mana occurs near the phase transition.
  }
  \label{fig:meanfield}
\end{figure}

As seen in Fig.~\ref{fig:meanfield}, the mana and magnetization are both identically zero in the paramagnetic phase and then jump to non-zero values just inside the ferromagnetic phase. Deep inside the ferromagnetic phase, the magnetization approaches its maximal value, and consequently, the mana approaches zero.

Additional analytic insight can be gained at large $q$. In this limit, numerics suggest that the mean field state is either a paramagnet or a strongly polarized ferromagnet with $\alpha \sim 1$. The energy per vertex is $\epsilon_p = -2 \cos \theta$ in the paramagnetic phase, while the energy per vertex is $\epsilon_f = -k \sin \theta$ in the ferromagnetic phase provided $\langle Z \rangle \approx 1 - O(1/q) $. So at infinite $q$, the transition is strongly first order and occurs at $\theta_c^{(\infty)}$ given by
\begin{equation}
    \cot \theta^{(\infty)}_c = k/2.
\end{equation}

To leading order in $1/q$, the $\alpha$ parameter in the ferromagnetic phase obeys
\begin{equation}
    \alpha = 1 - \frac{ \cot^2 \theta}{2qk^2}.
\end{equation}
So the system is already strongly polarized just inside the ferromagnetic phase. Accordingly, the mana per vertex of the mean field state is always of order one or less, much less than the maximal value of order $\ln q$.
We show this saturation in Fig.~\ref{fig:meanfield} bottom

\section{Discussion}\label{s:discussion}

In this manuscript we computed the mana of the ground states of some simple many-body systems,
and found that that magic is extensive.
For the $q = 3$ Potts model in 1d,
which has a second-order phase transition,
the mana density is an appreciable fraction of the upper bound $\frac 1 2 \ln 3$,
and reaches its peak at the phase transition.
For $q > 3$, where the transition should be well-described by a mean field theory,
we find that that the magic is still extensive, but much less than $\frac 1 2 \ln q$.

Measuring the mana of small, simply-connected subsystems
does not rule out the possibility that the magic is a local, lattice-scale effect,
and hence uninteresting from the point of view of either computation or gravity.
To check that possibility, we measured the ``connected component of mana''
on subsystems consisting of two widely-separated regions---that is, the degree to which the mana of the subsystem exceeded the mana of the two regions.
This connected component of mana displays a power-law decay,
indicating that the system's mana resides in large part in correlations
and cannot be removed by local unitaries.
When the two regions making up the subsystem have zero mana,
we additionally saw a ``sudden death of magic''.
This can be explained by the geometry of stabilizer states.

With the empirical behavior of the ground state magic established,
we gave a simple MERA picture for that behavior.
A MERA representation of a state is,
among other things,
a quantum circuit that prepares that state.
The structure of that circuit explains in some detail our results on the connected component of mana.
Additionally, by counting tensors in a region's past domain of dependence
we estimated the number of T-gates required to prepare that region.
This estimate heuristically explains the $\theta$-dependence and finite subsystem effects we see in the mana.
The fact that this estimate---which is (roughly) an upper bound on the number of non-Clifford rotations required,
since it comes from a specific circuit---largely matches the mana, which is a rough lower bound, indicates that recompilation can only slightly reduce the T gate count.
Much of the magic is therefore best understood as nonlocal---as residing in gates that affect many scales, including the largest.



We now discuss in more detail both the generality of our results and the motivations described in the introduction. Then we comment on possible new directions going forward and a number of outstanding questions.

\subsection{Extension to other models}

How well do our conclusions about the form of the magic generalize?
The overwhelming majority of our results are for one particular one-dimensional model,
the $\mathbb Z_3$ Potts model of Eq.~\ref{eq:ham},
and we measure magic using the mana (Eq.~\ref{eq:mana}),
which is only defined for odd prime Hilbert space dimensions.

But near the critical point our results follow from the MERA estimates of Sec.~\ref{s:mera};
for the two-point mana (Sec.~\ref{ss:results:twopoint}) we additionally have a phenomenological picture.
All of these pictures predict that the states should be magical
as measured by other, less convenient monotones,
and all should apply (\textit{mutatis mutandis}) to other models with continuous phase transitions.
Moreover our mean-field calculation (Sec.~\ref{s:meanfield})
sheds light on models with first-order phase transitions
(e.g. the $q \ge 4$ Potts model).

We therefore believe that the broad form of the mana of Fig.~\ref{fig:contiguous}
generalizes to models with continuous phase transitions.

\subsection{Implications for quantum simulation of field theories}
The results of Sec.~\ref{s:results},
especially Fig.~\ref{fig:contiguous},
suggest that magic state distillation will dominate the cost of preparing ground states of field theories.
If non-Clifford gates are much more time-consuming or otherwise expensive than Clifford gates,
then the difficulty of preparing the ground state of a field
will be controlled by the number of T gates required;
mana provides an (estimated) lower bound on the number of T gates required.

We find that for any $\theta \ne 0, \pi$ the ground state mana is extensive,
so the number of $T$ gates required is also extensive.
Since the number of gates required for state preparation is also extensive
(the quasi-MERAs of Sec.~\ref{ss:mera:nearcritical} give a concrete demonstration),
T gate cost will dominate the total cost of the calculation for any $\theta$ above some $\theta_{\text{crossover} }$,
the value of which depends on the details of the calculation and the difficulty of T gate application.

The MERA calculations of Sec.~\ref{s:mera} suggest that
a substantial fraction of the system's magic
could be removed by applying appropriate two-site unitaries.
If the MERA is known, these are simply the Hermitian conjugates of the blue square disentanglers;
if it is not, they can be found by optimizing the local basis that removes such local magic. 

We can therefore reduce the T gate cost of simulation
by working in the basis given by these unitaries,
at the cost of making the two-local Hamiltonian \eqref{eq:ham} three-local.

\subsection{AdS-CFT}
Tensor networks~\cite{Swingle:2009bg} and quantum error correction codes (QECC)~\cite{Almheiri:2014lwa} are intimately related to AdS/CFT.
They offer concrete geometric representations of the CFT entanglement pattern and aspects of the subregion duality, such as entanglement wedge reconstruction~\cite{Dong:2016eik}.
But holographic error correction codes typically have a flat entanglement spectrum,
which the CFT does not~\cite{CalLef2008}.
Because certain features of the entanglement spectrum are believed to be connected to gravity in the holographic theory~\cite{Dong:2016fnf} and other non-universal properties of certain CFTs, it is natural to ask what additional ingredients are necessary for our existing quantum error correction code models~\cite{pastawski_holographic_2015, Hayden:2016cfa} to reproduce them. 

We have attacked this question by constraining the circuits one might use to construct ground states of CFTs.
Encoding circuits for stabilizer code models of holographic QECC can be constructed using Clifford gates.
Because we need only a T gate to complete the universal gate set,
it must be the remaining necessary and sufficient element needed to reproduce the interesting features of the CFT entanglement spectrum.
Our quantitative results indicate that the CFT ground state can be almost as far as possible from a stabilizer state.
In addition, because magic is distributed non-locally,
our findings rule out the possibility
of modeling the ground state by a stabilizer circuit with a layer of T gates at the lattice scale.

This distribution of T gates is consistent with our intuition from holographic tensor networks.
The tensor network
(which we can understand as an encoding circuit, \`a la Sec.~\ref{s:mera})
should preserve the symmetries of the bulk, so the distribution of T gates should as well.
So at least for holographic CFTs the distribution of magic should be multiscale---as we in fact find.
Indeed, because the MERA picture captures the leading order behaviour of non-local magic, it is a strong indication that a multiscale structure like the MERA's hyperbolic geometry is close to the optimal form with which to distribute magic.

If we wish to reproduce the entanglement spectrum in QECC models, the existing network structure may not need significant reworking.
Instead, magic in the form of T-gates may be assigned locally to each tensor element of the QECC before they are wired together in a tensor network.
For instance, one might replace the 5-qubit encoding circuits used in the HaPPY constructions \cite{pastawski_holographic_2015}
with ones that prepare approximate quantum error correction codes,
in which some T gates are added to the original Clifford encoding circuit.
These codes have been shown to support the correct spectrum~\cite{Kim:2016wby}
and power law correlations~\cite{Hayden:2016cfa,CaoLackey}.


\subsection{The broad scope of magic studies}
While we only touch on a few specific aspects of magic and many-body physics in this work, one can tackle a far wider range of topics using magic.
In particular, the study of many-body magic is complementary to the work on optimizing circuits for quantum computation \cite{heyfron_efficient_2018,wang_resource-optimized_2020,shaw_quantum_2020}. That work seeks to find circuits that minimize the number of T gates used in some computation, whereas many-body magic lower-bounds that number.

Here is one example of the broader relevance of magic to many-body physics. Let us compare the circuits required to prepare the $3$-state Potts critical ground state in 1d and the toric code ground state in 2d. Both require a MERA circuit to produce their ground states---no finite depth circuit is sufficient. Hence, both systems possess long-range entanglement. One of the main results of this work is that the Potts ground state also requires long-range magic, i.e. magic at all scales in the MERA network. However, the toric code ground state is an exact stabilizer state and hence is non-magical. Moreover, any ground state in the same phase of matter has only short-range magic. This is because, by adiabatic continuity, any ground state in the phase is related by a short-depth circuit to the toric code ground state.\footnote{More precisely, quasi-adiabatic continuity can be used to construct a transformation using a Hamiltonian evolution where the Hamiltonian is local up the tails that decay with distance faster than any power.} Hence, all the magic can be removed by a short-depth circuit.

By characterizing magic in many-body systems, this work immediately prompts many questions.
For example, how can we sharpen this notion of ``nonlocal magic'', or magic ``residing in correlations''?
(One can define magic monotones for quantum channels \cite{seddon_quantifying_2019,wang_quantifying_2019,li_quantifying_2020,liu_operational_2020};
applying these measures to the MERA representation of a state may be a fruitful direction.)
What can we learn from a more careful treatment of finite subsystem effects?
Magic is important for state preparation in conformal field theories.
Can the amount and distribution of magic tell us anything about the bulk theory in AdS/CFT? 
For instance, the Renyi spectrum for disjoint subregions is believed to capture features of holographic conformal field theories~\cite{Dong:2016fnf}. Hence, non-local magic, like non-local entanglement, may be an important factor for identifying holographic CFTs, or other non-universal features. On a more aspirational note, can it help us evaluate the algorithmic hardness in simulating holographic CFTs both classically and quantum mechanically? Can it tell us anything about the complexity of the holographic dictionary\cite{Bouland:2019pvu}? Is there a ``magical Ryu-Takayanagi formula'' that relates magic in the CFT to bulk theory quantities?

Many technical matters also require attention. For example, the difficulty of computing mana scales exponentially with system size. We need a more efficient way to compute, or at least estimate mana for large systems. This may come from a better understanding of the distribution of Wigner coefficients, reducing the computational complexity using the symmetries of the system, and possibly a more tractable measure of magic---ideally one where we can calculate with difficulty polynomial in MPS bond dimension. Additionally, systems of interest frequently have symmetries, whether it is a $\mathbb Z_q$ ``spin flip'' like the Potts model, a $U(1)$ particle number symmetry, or an $SU(3)$ gauge symmetry. Such symmetries should provide additional structure for how magic is distributed in the state. For example, where does magic reside in a highly-entangled state in a given particle-number sector? 

Looking beyond these immediate technical questions, we are also hopeful that magic can be used as a diagnostic of the information structure of other many-body quantum effects. For example, can we use magic to distinguish different orders and phases in quantum many body problems, contrasting phases with short- and long-range magic? Also, what can magic tell us about the process of many-body quantum thermalization, and the difficulty of simulating hydrodynamical behavior?


\begin{acknowledgements}
  Our simulation and analysis scripts are available at \url{https://github.com/christopherdavidwhite/potts-gs-magic}; they use ITensors.jl v0.1.13 (\url{http://itensor.org/}).
  The plots were created in an IJulia (Jupyter) notebook available on request.
  
 We are grateful to Xin Wang, for many helpful early discussions about magic; 
 Brenden Roberts, for technical discussions about DMRG near the critical point and broadly insightful comments;
 Scott Lawrence, for discussions of quantum Monte Carlo and its relation to magic;
 Troy Sewell, for a number of insightful questions, and John Martyn, for many helpful conversations.

CDW is grateful to the U.S. Department of Energy (DOE), Office of Science, Office of Advanced Scientific Computing Research (ASCR) Quantum Computing Application Teams program, for support under fieldwork proposal number ERKJ347. CC and BGS acknowledge the support by the DOE Office of Science, Office of High Energy Physics, through the grant de-sc0019380. CC is also grateful for the support by the U.S. Department of Defense and NIST through the Hartree Postdoctoral Fellowship at QuICS.
\end{acknowledgements}

\bibliography{references}
\appendix

\section{Ground state search}\label{app:gs}

\begin{figure}[t]
  \begin{minipage}{0.45\textwidth}
    \includegraphics[width=\textwidth]{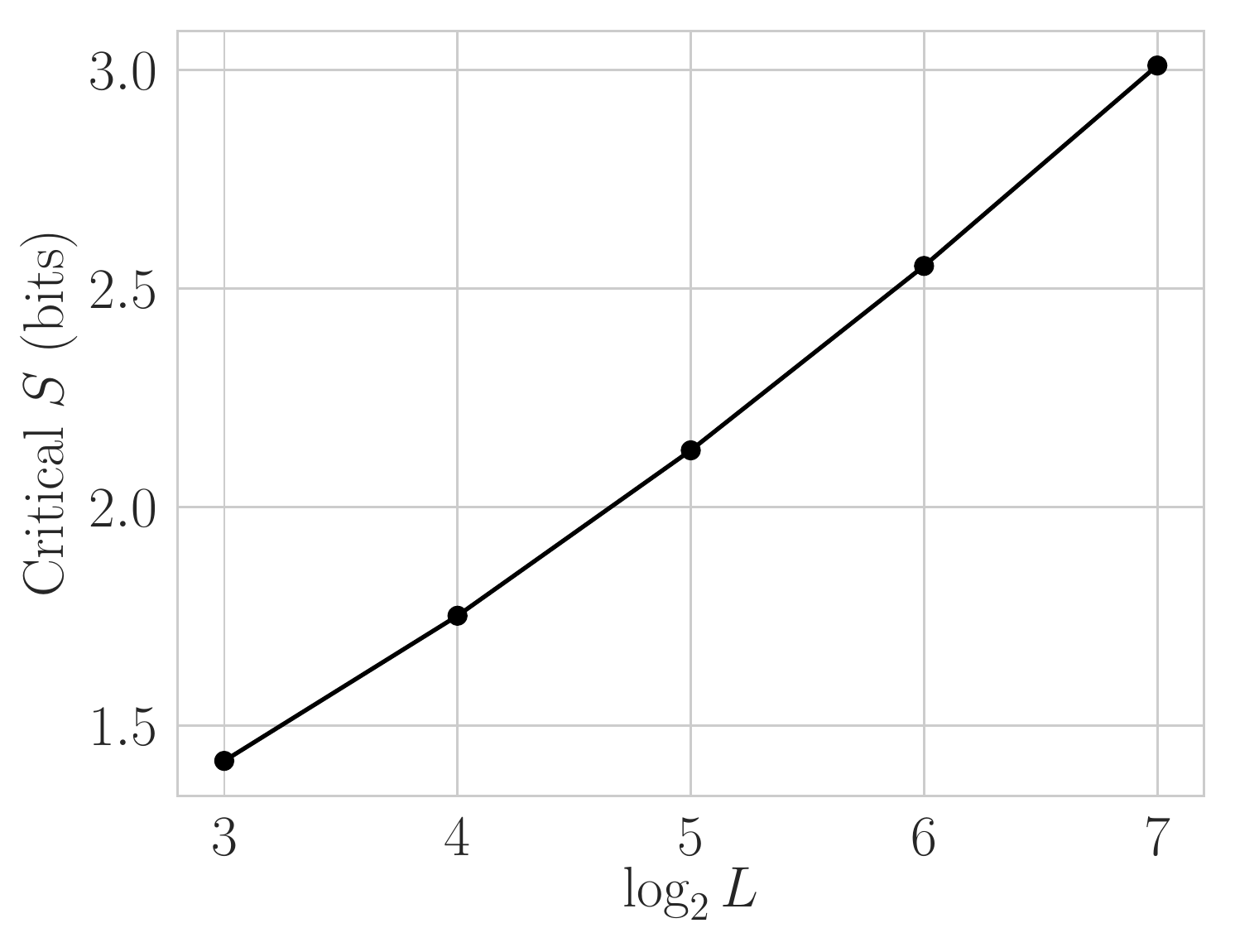}
  \end{minipage}
  \begin{minipage}{0.45\textwidth}
    \includegraphics[width=\textwidth]{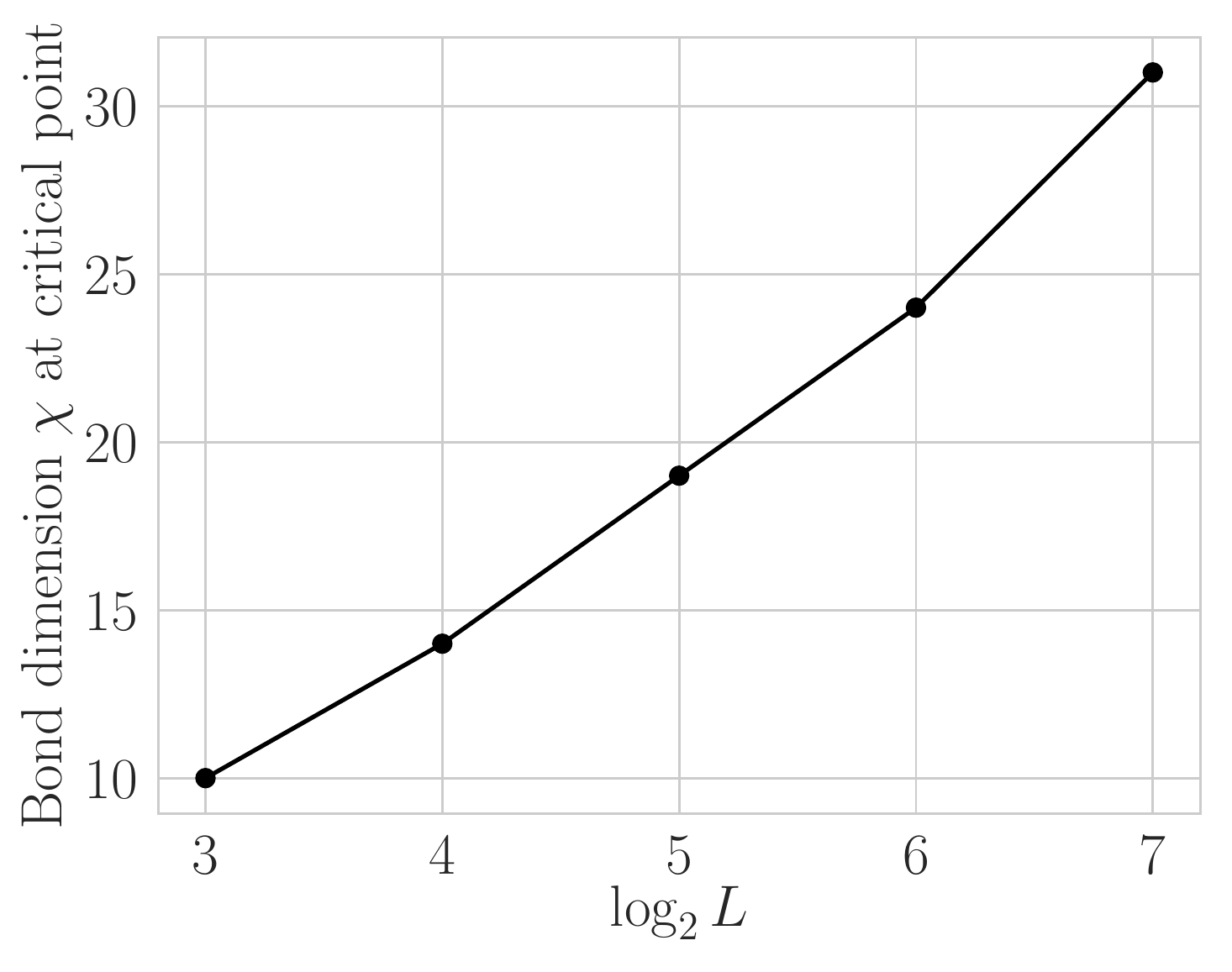}
  \end{minipage}
  \caption{\textbf{Entanglement entropy (top) and bond dimension (bottom)} at the critical point as a function of system size, $\lambda = 0$. }
  \label{fig:app:ent}
\end{figure}

\begin{figure}[t]
  \includegraphics[width=0.45\textwidth]{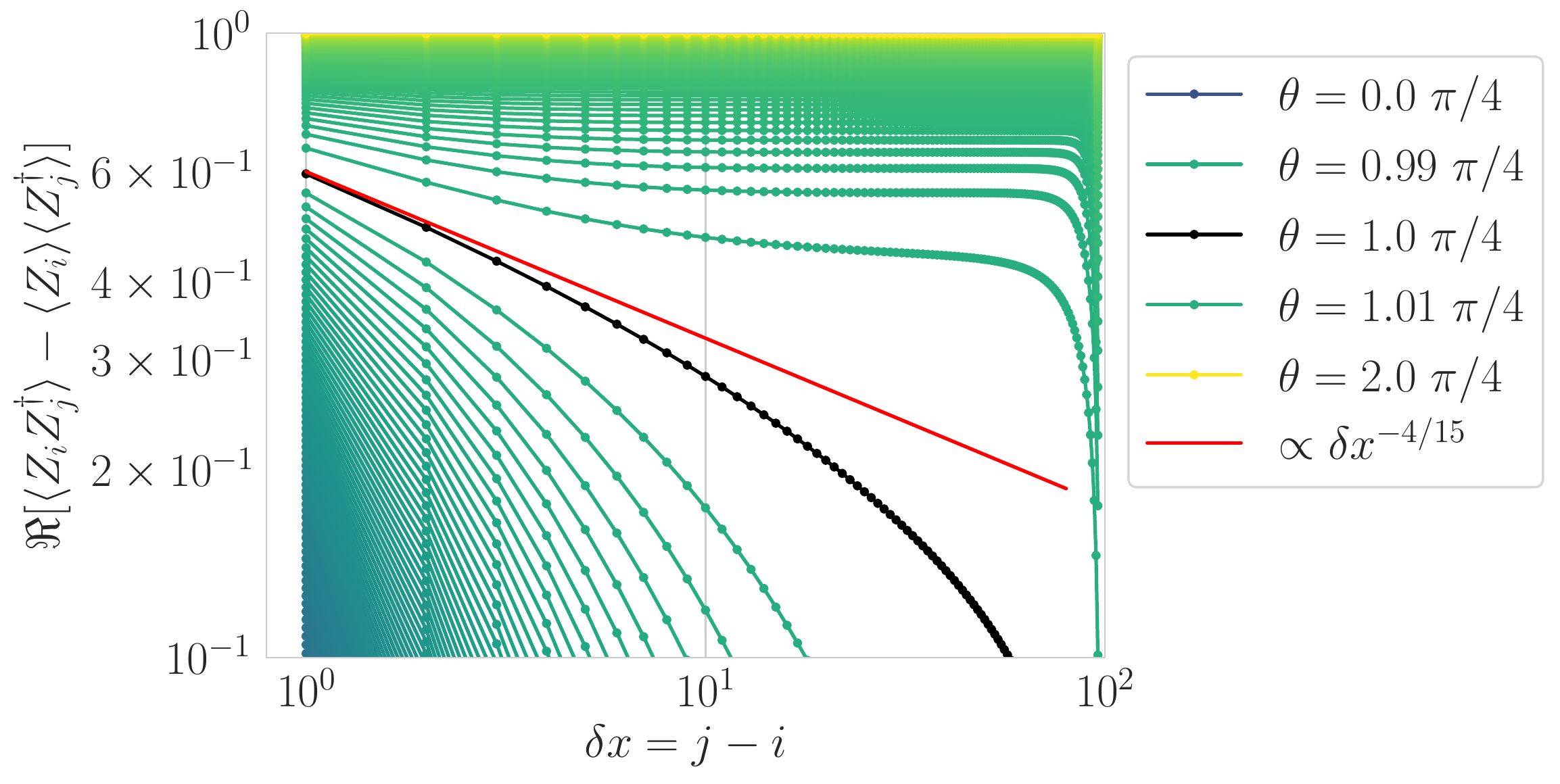}
  \caption{\textbf{Correlation functions} as a function of model parameter $\theta$ at $\lambda = 0$ for the cat state $\ket{\Omega}$.
    Our DMRG computations underestimate the long-range critical correlations.
  }
  \label{fig:app:corr-dx}
\end{figure}

We use DMRG to find the ground state of the 3-state Potts model 
\begin{equation*}
  \label{eq:ham}
  H = -\sin \theta \sum_j [Z_j^\dagger Z_{j+1} + h.c.] - \cos \theta \sum_j [X^\dagger_j + X_j]\;.
\end{equation*}
We take both an SVD truncation cutoff and DMRG energy convergence threshold $10^{-7}$. 
We find that the entanglement entropy at the critical point is logarithmic in system size and not too large for $N = 128$ , and the bond dimension is always tractable (Fig.\ref{fig:app:ent}).
The clean separation between $\theta < \pi / 4$, $\theta = \pi / 4$, and $\theta > \pi / 4$ in the correlations (Fig.~\ref{fig:app:corr-dx}) and the two-point mana (Fig.~\ref{fig:twosite-mera})
empirically confirm that we are capturing some of the critical behavior, at least for $N = 128$ sites.
But in Fig.~\ref{fig:app:corr-dx} we see that at the critical $\theta = \pi / 4$
the correlations ultimately decay faster than a power law,
indicating that high-precision determination of (e.g.) the large-$\delta x$ behavior of the connected component of mana would require more effort.

\begin{figure}
  
  \begin{minipage}{0.40\textwidth}
    \includegraphics[width=\textwidth]{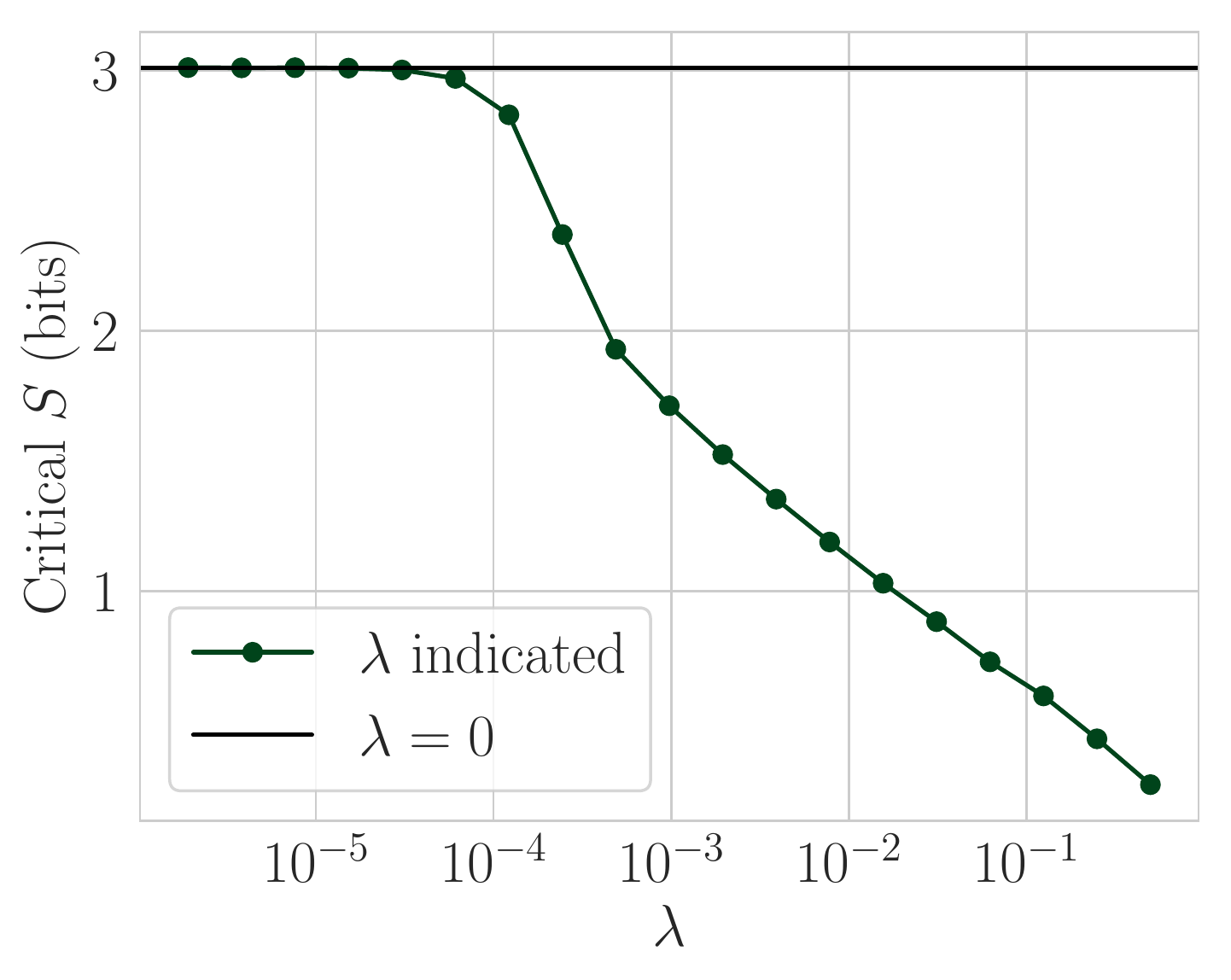}
  \end{minipage}
  \begin{minipage}{0.40\textwidth}
    \includegraphics[width=\textwidth]{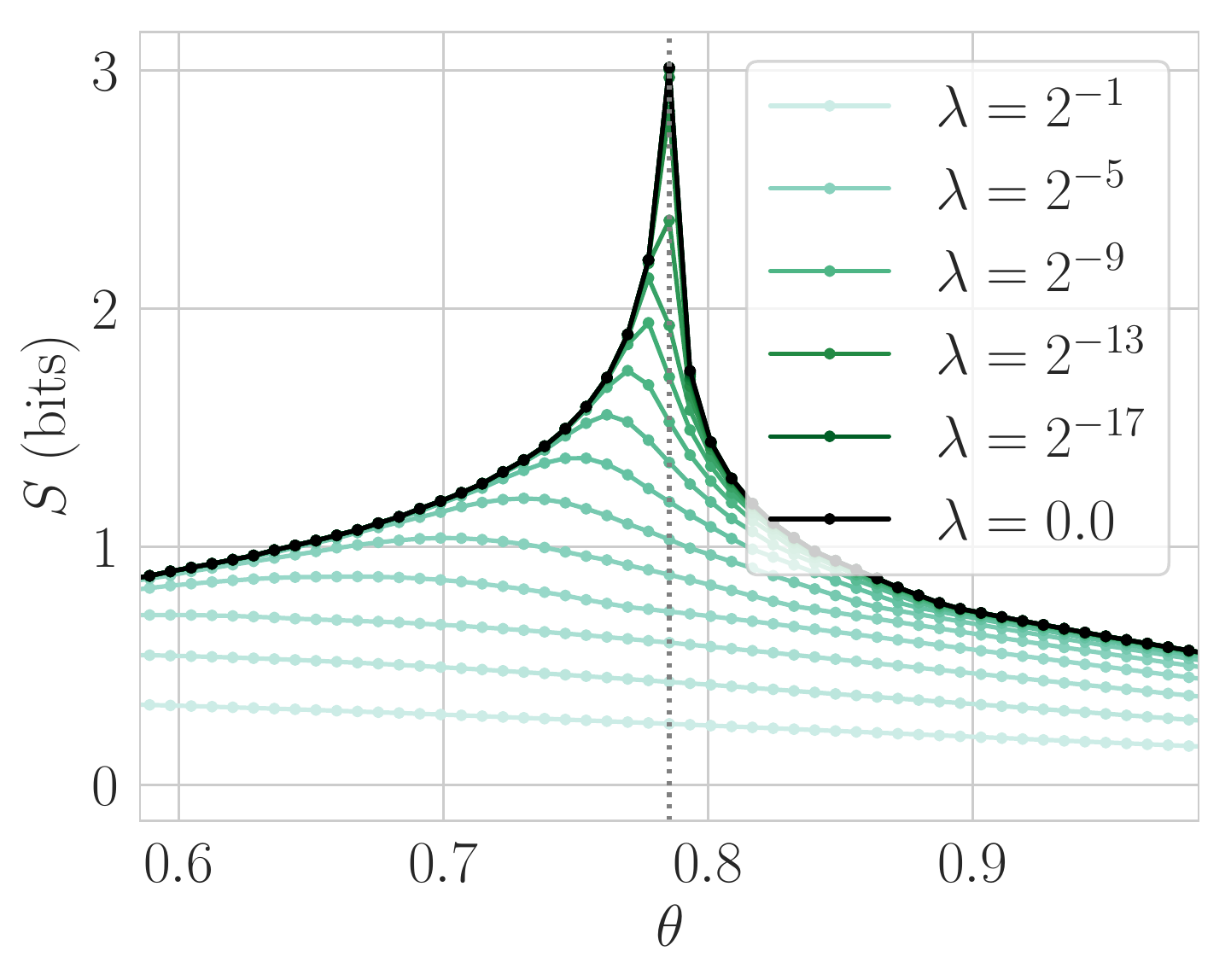}
  \end{minipage}
  \begin{minipage}{0.40\textwidth}
    \includegraphics[width=\textwidth]{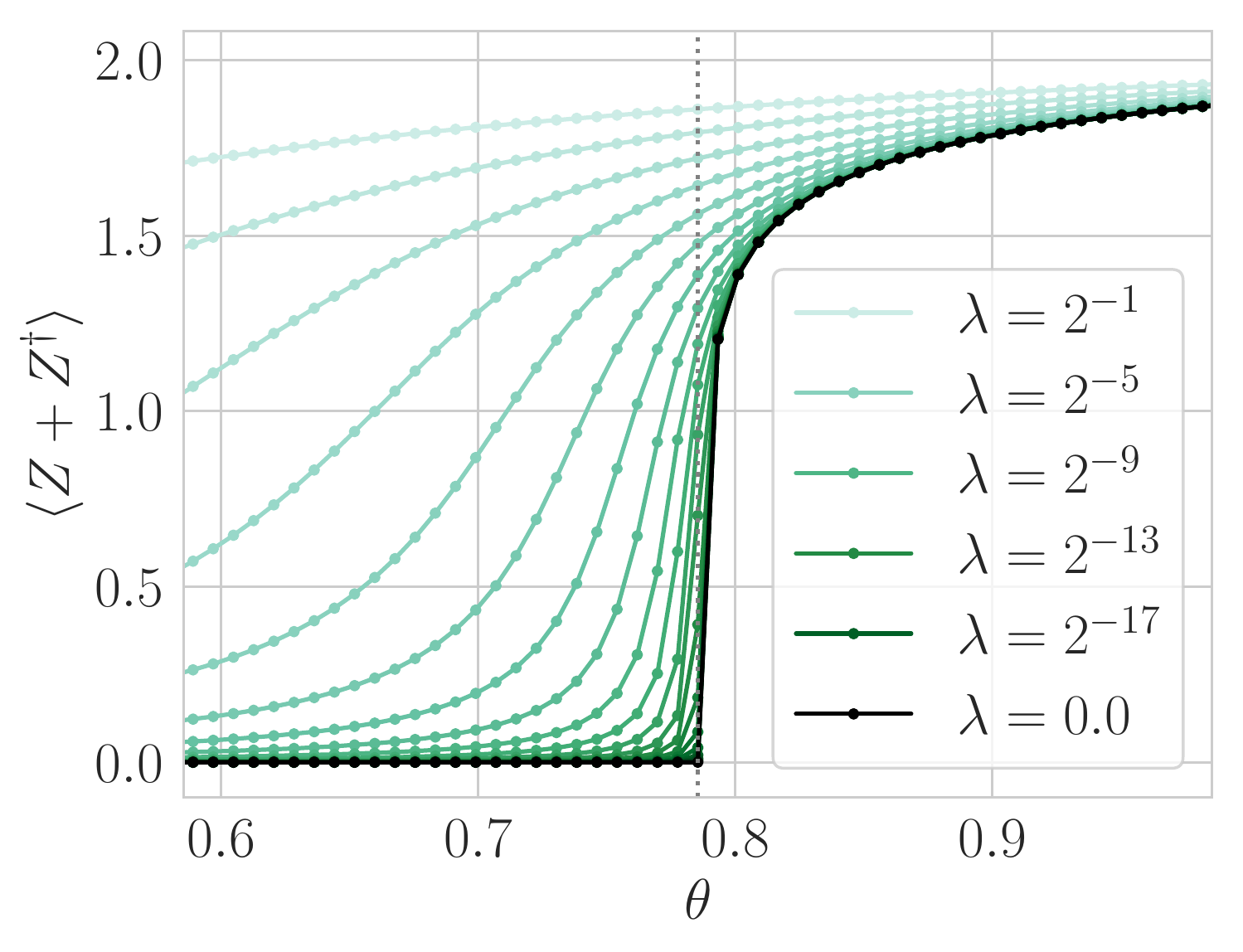}
  \end{minipage}
  \caption{\textbf{Convergence in external field $\lambda$:}
    von Neumann entanglement entropy (at the critical point $\theta = \pi/4$, \textbf{top}, and as a function of $\theta$,  \textbf{middle}) and $\expct{Z + Z^\dagger}$ (\textbf{bottom}) in the symmetry-broken ground state $\ket {\Omega_0}$.
    The dashed line shows the critical $\theta = \pi/4$.
  }
  \label{fig:app:long}
\end{figure}

The system's threefold degeneracy in the ferromagnetic phase requires some thought.
As discussed in Sec.\ref{s:method}, we
first find a symmetry-broken ground state $\ket{\Omega_0}$ by biasing the initial state
and then 
construct the cat ground state
\begin{equation}
  \ket \Omega = \sum_n \Big[\prod_j X_j \Big]^n \ket {\Omega_0}\;.
\end{equation}
We check that this bias is consistent with the result of adding a small longitudinal field, for 
\begin{align*}
  H &= -\sin \theta \sum_j [Z_j^\dagger Z_{j+1} + h.c.]\\
  &\quad - \cos \theta \sum_j [X^\dagger_j + X_j] - \lambda \sum[Z_j + Z_j^\dagger]\;,
\end{align*}
does not change our results substantially
(see Fig.~\ref{fig:app:long} for entanglement entropy and order parameter).

\subsection{Justification for the cat state}

\subsubsection{Finite systems}
For a finite system of length $N$, the ground state is not strictly threefold degenerate.
the global shift $\prod_j X_j$ shows up at $N$th order in perturbation theory, and splits the energies by some amount exponentially small in system size.
The true ground state, then, is the cat state $\ket \Omega$.

(DMRG does not see this energy difference, which should be smaller even than machine precision;
also, ``intermediate states'' reachable by $\prod{j \in J \subsetneq 1\dots N} X_j$ have an $O(1)$ energy penalty.)

\subsubsection{Long systems and duality}

For infinite systems, we can justify the cat state by appealing to the model's duality mapping and to physical intuition.
In this case we understand the cat state as reducing finite-system-size effects.

First, recall the duality mapping. For convenience relabel the coupling constants
\begin{equation}
  J = \sin \theta \;,\qquad h = \cos \theta
\end{equation}
so the Hamiltonian is
\begin{equation}
  H = - J \sum_j [Z_j^\dagger Z_{j+1} + h.c.] - h \sum_j [X^\dagger_j + X_j]\;. 
\end{equation}
Map
\begin{align}
  \tilde X_j &:= Z_j Z_{j+1}^\dagger &\text{measures domain wall at bond $j$}\\
  \tilde Z_j &:= \prod_{k > j} X_k &\text{increments domain wall at bond $j$}
\end{align}
Now check that $\tilde X_j, \tilde Z_l$ multiply like clock and shift operators:
\begin{enumerate}
\item $j = l$:
  \begin{align*}
    \tilde X_j \tilde Z_j &= \left[Z_j Z_{j+1}^\dagger \right]\left[\prod_{k > j} X_k \right]\\
    &= \omega^{-1}\left[\prod_{k > j} X_k \right]\left[Z_j^\dagger Z_{j+1} \right] = \omega^{-1} \tilde Z_j\tilde X_j
  \end{align*}
  because $\tilde X_j, \tilde Z_j$ overlap only at site $j$.
\item $j < l$
  \begin{equation}
   \tilde  X_j\tilde  Z_l = \left[Z_j^\dagger Z_{j+1} \right]\left[\prod_{k > l} X_k \right] = \tilde Z_l \tilde X_j
 \end{equation}
 because  $\tilde X_j, \tilde Z_l$ do not overlap.
\item $l < j$
  \begin{equation}
   \tilde  X_j\tilde  Z_l = \left[Z_j^\dagger Z_{j+1} \right]\left[\prod_{k > l} X_k \right] = \tilde Z_l \tilde X_j
 \end{equation}
 because $Z_j^\dagger Z_{j+1} X_j X_{j+1} =  X_j X_{j+1} Z_j^\dagger Z_{j+1}$.
\end{enumerate}

Invert the mapping:
\begin{align*}
  Z_1Z_j^\dagger &= \prod_{1 \le k < j} \tilde X_k\\
  X_j &= Z_{j-1} Z_j^\dagger
\end{align*}
These are only defined for $j=2$.
The $Z_1 Z_j^\dagger$ construction is easier to understand once we

With the duality mapping defined on operators, we can map states.
Take a $\tilde X_j$ eigenstate
\begin{equation}
  \ket {\mathbf n} = \ket{n_1 \dots n_{N-1} }\;.
\end{equation}
This has
\begin{equation}
  Z_1Z_j^\dagger = \omega^{\sum_{1 \le k < j} n_k}\;.
\end{equation}
This is sensible.
Specifying $\tilde X_j$ specifies domain walls,
but not a site-1 starting state,
so it specifies only $Z$ correlations.
There is an ambiguity in the state:
we have to pick something for site $1$.

The natural way to resolve the ambiguity in the mapping is to fix
\begin{equation}
  Z_1 \ket {\mathbf n} = 0\;.
\end{equation}
But we want our numerical finite-size ground state to reflect as closely as possible the properties of the infinite-system ground state.
Consider it therefore as a subsystem of a much larger system.
Pin the left site of the much larger subsystem.
By the time we reach the start of the subsystem, we will have seen many domain walls---in fact, a superposition of different numbers of domain walls.\footnote{
  In effect we consider the MPS representation of the infinite-system ground state \emph{before} duality mapping as a Markov chain.
}
The left edge of the subsystem is therefore equally likely to be in any of the three states.
To model this precisely,
we should take a statistical mixture of the three states,
but we take the cat state $\ket{\Omega}$ for convenience.

\begin{figure}[t]
  \begin{minipage}{0.45\textwidth}
  \includegraphics[width=\textwidth]{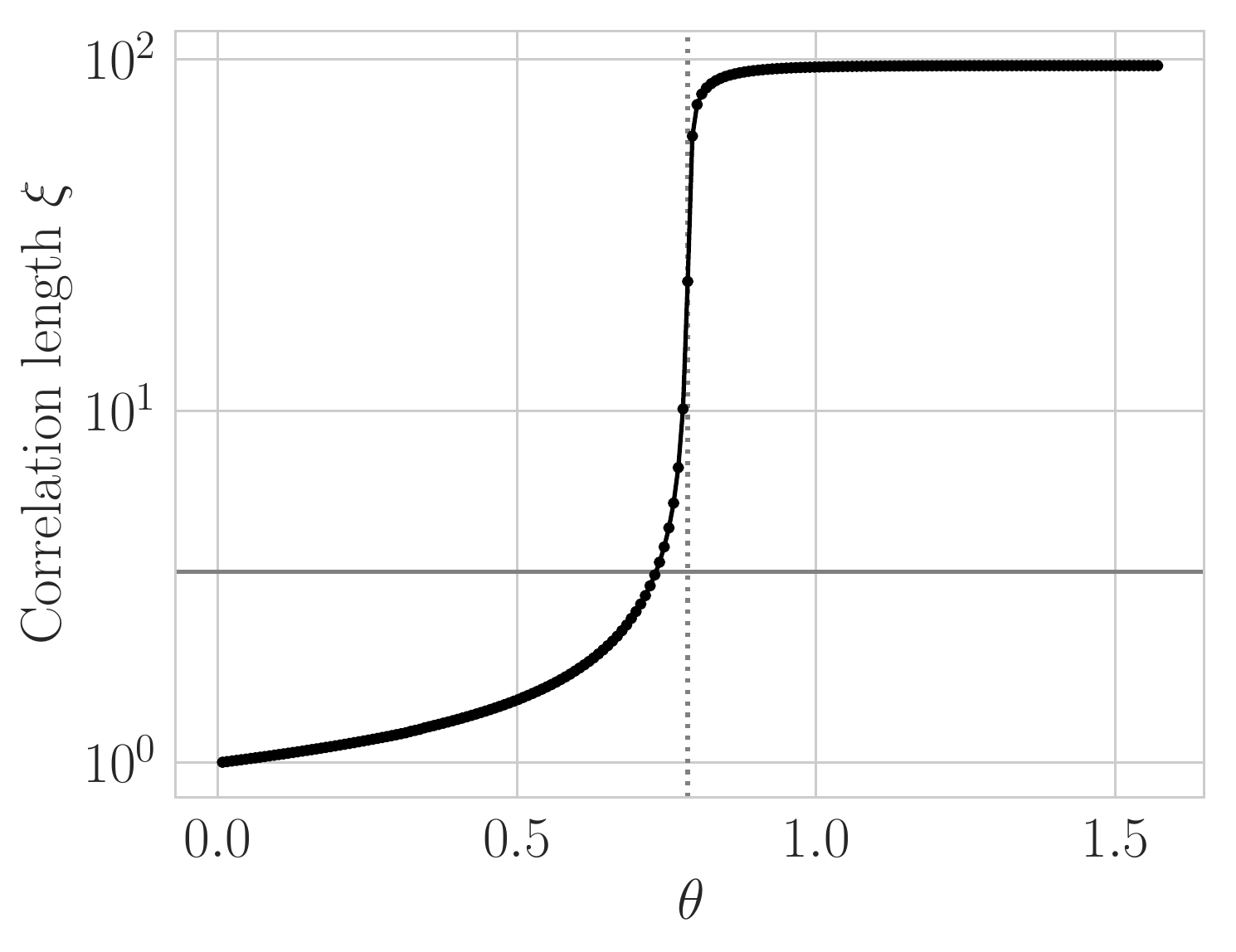}
  \end{minipage}


  \caption{\textbf{Correlation lengths}
    extracted from the $ZZ$ correlation function Eq.~\eqref{eq:app:corr:def}
    (cf Fig.~\ref{fig:app:corr-dx})
    via Eq.~\eqref{eq:app:corr:xi-def}
    for the cat state $\ket{\Omega}$.
    The vertical grey dotted line 
    marks the critical $\theta_c = \pi/4$;
    the horizontal grey solid line marks $\ell/2$ for $\ell = 7$,
    the largest subsystem whose mana we compute.
  }
  \label{fig:app:corr-theta}
\end{figure}

\section{Correlation lengths}\label{app:corr}

Fig.~\ref{fig:app:corr-dx} shows
\begin{equation}
  \label{eq:app:corr:def}
  C_{ij}(\theta) := \Re \left[\expct{Z_i Z^\dagger_j} - \expct{Z_i}\langle Z^\dagger_j\rangle \right]
\end{equation}
for $i = L/4 = 32$.
We define correlation lengths
\begin{equation}
  \label{eq:app:corr:xi-def}
  \xi = C_{i,i+1}^{-1} \sum_{j > i} C_{ij}
\end{equation}
and plot them in Fig.~\ref{fig:app:corr-theta}.
\end{document}